\documentclass[fleqn,useAMS,usenatbib]{mn2e}

\usepackage{graphicx}
\usepackage{amsmath}
\usepackage{mathrsfs}

\makeatletter

\def\compoundrel#1\over#2{\mathpalette\compoundreL{{#1}\over{#2}}}
\def\compoundreL#1#2{\compoundREL#1#2}
\def\compoundREL#1#2\over#3{\mathrel
      {\vcenter{\hbox{$\m@th\buildrel{#1#2}\over{#1#3}$}}}}
\makeatother

\newcommand{\msun}{M_{\odot}}

\title[Dark-matter halos and the $M$--$\sigma$ relation]
  {Dark-matter halos and the {\boldmath$M$--$\sigma$} relation for
    supermassive black holes}
\author[A.C.~Larkin and D.E.~McLaughlin]
  {Adam C. Larkin\thanks{E-mail: a.larkin@keele.ac.uk} and
    Dean E. McLaughlin\thanks{E-mail: d.e.mclaughlin@keele.ac.uk}
    \\
    Astrophysics Group, Lennard-Jones Laboratories, Keele University,
    Keele, Staffordshire, ST5 5BG, UK} 

\defcitealias{NFW97}{NFW}

\begin{document}

\date{\today}

\pagerange{\pageref{firstpage}--\pageref{lastpage}} \pubyear{2015}

\maketitle 

\label{firstpage}

\begin{abstract}
We develop models of two-component spherical galaxies to establish
scaling relations linking the properties of spheroids at
$z=0$ (total stellar masses, effective radii $R_e$ and velocity
dispersions within $R_e$) to the properties of their dark-matter halos
at both $z=0$ and higher redshifts. Our main motivation is the widely
accepted idea that the accretion-driven growth of supermassive
black holes (SMBHs) in protogalaxies is limited by quasar-mode
feedback and gas blow-out. The SMBH masses, 
$M_{\rm{BH}}$, should then be connected to the dark-matter potential
wells at the redshift $z_{\rm{qso}}$ of the blow-out.
We specifically consider the example of a power-law dependence
on the maximum circular speed in a protogalactic dark-matter halo:
$M_{\rm{BH}}\propto V^4_{\rm{d,pk}}$, as could be expected if quasar-mode
feedback were momentum-driven. For halos with a given $V_{\rm{d,pk}}$
at a given $z_{\rm{qso}}\ge 0$, our model scaling relations give a typical
stellar velocity dispersion $\sigma_{\rm{ap}}(R_e)$ at $z=0$. Thus,
they transform a theoretical ``$M_{\rm{BH}}$–-$V_{\rm{d,pk}}$ relation''
into a prediction for an observable
$M_{\rm{BH}}$-–$\sigma_{\rm{ap}}(R_e)$ relation. We find the latter to be
distinctly non-linear in log–-log space. Its shape depends on the
generic redshift-evolution of halos in a {$\Lambda$}CDM cosmology and
the systematic variation of stellar-to-dark matter mass fraction at
$z=0$, in addition to any assumptions about the physics underlying the
$M_{\rm{BH}}$-–$V_{\rm{d,pk}}$ relation. Despite some clear
limitations of the form we use for $M_{\rm{BH}}$ versus $V_{\rm{d,pk}}$,
and even though we do not include any SMBH growth through dry mergers
at low redshift, our results for
$M_{\rm{BH}}$--$\sigma_{\rm{ap}}(R_e)$ compare well to data for local
early types if we take $z_{\rm{qso}}\sim 2$-–$4$.
\end{abstract}


\section{Introduction}
\label{sec:intro}

The masses $M_{\rm BH}$ of supermassive black holes (SMBHs) at the
centres of normal early-type galaxies and bulges correlate with
various global properties of the stellar spheroids---see
\citet{kho13} for a comprehensive review. 
The strongest relationships include one between
$M_{\rm BH}$ and the bulge mass $M_{\rm bulge}$
(either stellar or dynamical, depending on the author: e.g.,
\citealt{magetal98,mh03,hnr04,mcconnellma});
a scaling of $M_{\rm BH}$ with the (aperture) stellar velocity
dispersion $\sigma_{\rm ap}$ averaged inside some fraction of the
effective radius $R_e$ of the bulge
($M_{\rm BH}\sim \sigma_{\rm ap}^{4{\mbox{--}}5}$ if fitted with a
single power law:
\citealt{ferrarese00,gebhardt00,fandf,mcconnellma});
and a fundamental-plane dependence of $M_{\rm BH}$ on a
combination of either $M_{\rm bulge}$ and $\sigma_{\rm ap}$ or
$\sigma_{\rm ap}$ and $R_e$ (\citealt{hop07b,hop07c}).
Whether any one correlation is more fundamental than the
others is something of an open question, but collectively they are
interpreted as evidence for co-evolution between SMBHs and their host
galaxies.

This co-evolution likely involved self-regulated feedback in
general. Most of the SMBH mass in large galaxies is grown in a
quasar phase of Eddington-rate accretion \citep{yu02}, driven by a
rapid succession of gas-rich mergers at high redshift. Such accretion
deposits significant momentum and energy back into the protogalactic
gas supply, which can lead to a blow-out that
stops further accretion onto the SMBH. In this context, the empirical
correlation between $M_{\rm BH}$ and $\sigma_{\rm ap}$ takes on
particular importance, as the stellar velocity dispersion
should reflect the depth of the potential well from which SMBH
feedback had to expel the protogalactic gas. Cosmological
simulations of galaxy formation now routinely include
prescriptions for the quenching of Eddington-rate accretion by
``quasar-mode'' feedback, with free parameters that are tuned to give
good fits to the SMBH $M$--$\sigma$ relation at $z=0$.

However, it is not clear in detail how the stellar velocity
dispersions in normal galaxies at $z=0$ relate to the protogalactic
potential wells when any putative blow-out occurred and the main phase
of accretion-driven SMBH growth came to an end.
For most systems, this
was presumably around
$z\sim 2$--3, when quasar activity
in the Universe was at its peak \citep{rich06,hop07a}.
The potential wells in question were
dominated by dark matter, and a general method is lacking to
connect the stellar $\sigma_{\rm ap}$ in spheroids
to the properties of their dark-matter halos, not only at
$z=0$ but at higher redshift as well.
Moreover, it is not necessarily obvious what specific property (or
properties) of dark-matter halos provides the key measure of
potential-well depth in the context of a condition for
accretion-driven blow-out. Different simulations of galaxy and SMBH
co-evolution with different recipes for quasar-mode feedback appear
equally able (with appropriate tuning of their free parameters) to
reproduce the observed $M$--$\sigma$ relation.

Our main goal in this paper is to address the first part of this
problem. We develop ``mean-trend'' scaling relations between
the average stellar properties (total masses, effective radii and
aperture velocity dispersions) and the dark-matter halos (virial
masses and radii, density profiles and circular-speed curves) of
two-component spherical galaxies. These scalings are constrained by
some data for a representative sample of local early-type galaxies,
and by the properties of dark-matter halos at $z=0$ in cosmological
simulations. We then include an analytical approximation to the
mass and potential-well growth histories of simulated dark-matter
halos, in order to connect the stellar properties at $z=0$ to halo
properties at $z>0$. We ultimately use these results to illustrate how
one particularly simple analytical expression, which gives a critical
SMBH mass for protogalactic blow-out directly in terms of the
dark-matter potential well at quasar redshifts, translates to a
relation between SMBH mass and stellar velocity dispersion at $z=0$.

\subsection{SMBH masses and halo circular speeds}
\label{subsec:mbhvhalo}

Under the assumption (which we discuss just below) that accretion
feedback is momentum-conserving and takes the form of a spherical
shell driven outwards by an SMBH wind with momentum flux
$dp_{\rm wind}/dt=L_{\rm Edd}/c$,
\citet{mcq12} derive a minimum SMBH mass sufficient to expel an
initially static and virialised gaseous medium from any
protogalaxy consisting of dark matter and gas only.
This critical mass is approximately
\begin{align}
M_{\rm BH} & ~\simeq~
  \frac{f_0 \kappa}{\pi G^2}\frac{V^{4}_{\rm{d,pk}}}{4}  
\notag \\
  & ~\simeq~ 1.14\times10^{8}\,M_\odot\,
         \left(\frac{f_0}{0.2}\right)
         \left(\frac{V_{\rm{d,pk}}}{200~{\rm km s^{-1}}}\right)^{4}
    ~ ,
\label{eq:msig}
\end{align}
where $\kappa$ is the Thomson-scattering opacity and
$f_0$ is the (spatially constant) gas-to-dark matter mass fraction in
the protogalaxy. The velocity scale $V_{\rm d,pk}$ refers to the
{\it peak} value of the circular speed
$V_{\rm d}^2(r)\equiv GM_{\rm d}(r)/r$ in a dark-matter halo with mass
profile $M_{\rm d}(r)$. Equation (\ref{eq:msig}) holds for any
form of the mass profile, just so long as the associated
circular-speed curve has a single, global maximum---as all realistic
descriptions of the halos formed in cosmological $N$-body simulations
do. Defining a characteristic (dark-matter) velocity
dispersion as $\sigma_0\equiv V_{\rm d,pk}/\sqrt{2}$ turns
equation (\ref{eq:msig}) into a critical $M_{\rm BH}$--$\sigma_0$
relation, which is formally the same as that obtained by
\citet{king03,king05}, and similar to the earlier result of
\citet{fabian99}, for momentum-driven blow-out from a singular
isothermal sphere.

\begin{figure}
\begin{center}
\includegraphics[width=83mm]{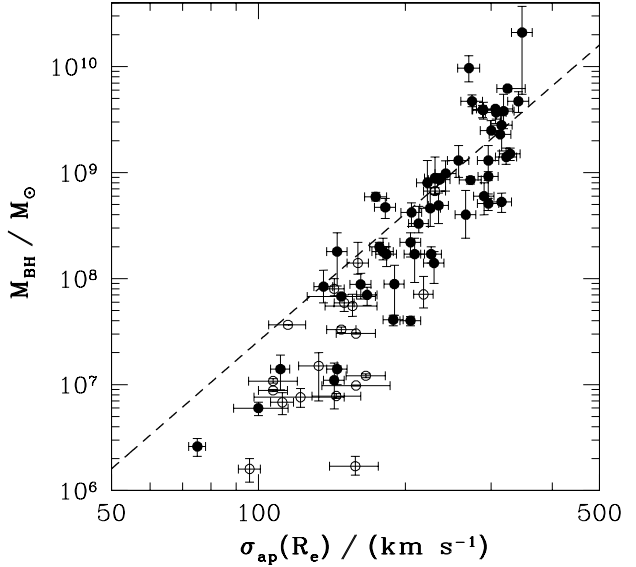}
\end{center}
\caption{SMBH mass versus stellar velocity dispersion averaged over an
  effective radius. Data are from the compilation of
  \citet{mcconnellma} for 53 E or S0 galaxies (filled circles) and
  19 bulges in late Hubble types (open circles). The dashed line is 
  equation (\ref{eq:msig}) with a protogalactic gas-to-dark matter
  fraction $f_0=0.18$ and
  $V_{\rm{d,pk}} \equiv \sqrt2\,\sigma_{\rm{ap}}(R_e)$ for all galaxies.
  Improving upon this poorly-justified association between the
  characteristic stellar and dark-matter velocities in early-type
  galaxies is one of the goals of this paper.}
\label{fig:simple}
\end{figure}

This critical mass is based on the simplified description given by
\citet{kp03} of a Compton-thick wind resulting from 
accretion at or above the Eddington rate onto an SMBH. In particular,
their analysis provides the assumption that the momentum flux in the
SMBH wind is simply $L_{\rm Edd}/c$ (with no pre-factor).\footnotemark
\footnotetext{Having $dp_{\rm wind}/dt=L_{\rm Edd}/c$, rather
  than $\propto\!L_{\rm Edd}/c$ but much less, implies high
  wind speeds of up to $\sim\!0.1\,c$ \citep{king10}.
  Such ``ultrafast outflows'' are observed in
  many local active galactic nuclei and low-redshift
  quasars accreting at or near their Eddington rates (e.g.,
  \citealt{pounds03,reeves03,tombesi10,tombesi11}).}
The wind from an SMBH with mass greater than that in equation
(\ref{eq:msig}) will then supply an outwards force (i.e.,
$L_{\rm Edd}/c=4\pi G M_{\rm BH}/\kappa$) on a thin, radiative shell
of swept-up ambient gas that exceeds the gravitational
attraction of dark matter behind the shell (maximum force
$f_0V_{\rm d,pk}^4/G$ if the gas was initially virialised),
{\it everywhere} in the halo. It is a
condition for the clearing of all gas to beyond the virial
radius of any non-isothermal halo.

Equation (\ref{eq:msig}) has limitations. Most notably, the
protogalactic outflows driven by SMBH winds are in fact
expected to become energy-driven (non-radiative) after
an initial radiative phase \citep{zub12,mcq13}. This may
\citep{silk98,mcq13} or may not \citep{zub14} change the functional
dependence of a critical $M_{\rm BH}$ for blow-out on the
dark-matter $V_{\rm d,pk}$ or any other characteristic halo velocity
scale. Beyond this, the equation also assumes a wind moving into an
initially static ambient medium, ignoring the cosmological
infall of gas and an additional, confining ram pressure that comes
with hierarchical (proto-)galaxy formation \citep{costa14}.
It also neglects the presence of any stars in protogalaxies, which
could contribute both to the feedback driving gaseous outflows
(e.g., \citealt{murray,power}) {\it and} to the gravity containing
them. (The assumptions of spherical symmetry and a
smooth ambient medium are not fatal flaws; see \citealt{zub14}).

However, it is not our intention here to improve equation
(\ref{eq:msig}). Rather, we aim primarily to establish a method
by which halo properties at $z>0$ in relations {\it such as} equation 
(\ref{eq:msig}) can be related to the average
properties of stellar spheroids at $z=0$. By doing this, we hope to
understand better how expected relationships between SMBH masses and
protogalactic dark-matter halos are reflected in the observed
$M$--$\sigma$ relation particularly.
Equation (\ref{eq:msig}) is a good test case because it is
simple and transparent but still contains enough relevant feedback
physics to be interesting, even with the caveats mentioned
above. It is also the only such relation we know of, which
{\it does not assume that dark-matter halos are singular isothermal
  spheres}.

\subsection{Halo circular speeds and stellar velocity dispersions}
\label{subsec:transform}

As a point of reference, Figure \ref{fig:simple} shows SMBH mass
against the stellar velocity dispersion $\sigma_{\rm ap}(R_e)$ within
an aperture equal to the stellar effective radius, for 
galaxies and bulges in the compilation of \citet{mcconnellma}. The
dashed line shows equation 
(\ref{eq:msig}) evaluated with a gas-to-dark matter mass ratio of
$f_0=0.18$ (the cosmic average; \citealt{planck14}) for all
protogalaxies at the time of blow-out, and with the naive substitution
$V_{\rm d,pk}\equiv \sqrt{2}\,\sigma_{\rm ap}(R_e)$ for all spheroids at
$z=0$. The proximity of this line to the data---first
emphasised by \citet{king03,king05}, who assumed isothermal
halos---encourages taking seriously the 
basic physical ideas behind equation (\ref{eq:msig}), even though (as
discussed above) some details must be incorrect at some level.

However, 
setting $V_{\rm d,pk}=\sqrt{2}\,\sigma_{\rm ap}(R_e)$ is problematic.
A $\sqrt{2}$-proportionality between circular speed and velocity
dispersion is appropriate only for isothermal spheres,
which real dark-matter halos are not.
A dark-matter velocity dispersion can be equated to a stellar velocity
dispersion only if the dark matter and the stars have the same spatial
distribution, which is not true of real galaxies.
And $V_{\rm d,pk}$ in equation (\ref{eq:msig}) refers to a
protogalactic halo, which will have grown significantly since
the quasar epoch at $z\sim2$--3.

In \S\ref{sec:modingred}, we gather results from the
literature that we need in order to address these issues. In
\S\ref{sec:modresults}, we combine them to constrain simple models of
spherical, two-component galaxies, focussing on
scaling relations between the stellar and dark-matter
properties at $z=0$. This is done without any reference to black
holes, and the scalings should be of use beyond
applications to SMBH correlations. In \S\ref{sec:msigma}, we
make a new, more rigorous comparison of equation (\ref{eq:msig}) to
the SMBH $M$--$\sigma$ data (compare Figure \ref{fig:mbhsigma} below
to Figure \ref{fig:simple}). Our work could in principle be used to
explore the consequences of SMBH--halo relations like equation
(\ref{eq:msig}) for other SMBH--bulge correlations as well, but we do
not pursue these here. In \S\ref{sec:summary}, we summarise the
paper. 

\section{Model Ingredients}
\label{sec:modingred}

Equation (\ref{eq:msig}) incorporates an assumption that gas traced
the dark matter in {\it protogalaxies} before being blown out by
quasar-mode accretion feedback at high redshift. However,
it does not
make any assumptions about the detailed structure of dark-matter
halos at any epoch, and it neither requires nor implies that mass
follows light in galaxies at $z=0$.

In this Section, we collect together analytical
expressions from the literature for the (different) stellar
and dark-matter mass profiles in galaxies, and for some key structural
parameters of dark matter halos and their evolution in $\Lambda$CDM
simulations of structure formation. We use these to obtain our new
results in \S\ref{sec:modresults} and \S\ref{sec:msigma}. Some of
these expressions from the literature, and all of the scaling
relations we ultimately derive, represent average trends that can have
significant scatter around them. We do not attempt in this paper to
analyse such scatter or to predict the net scatter
around any scaling that comes from combining others.

This Section and \S\ref{sec:modresults} do not rely on any ideas
about black hole accretion feedback or SMBH--bulge correlations. We
focus repeatedly on the peak circular speed $V_{\rm d,pk}$ in 
dark-matter halos, because that is what appears in equation
(\ref{eq:msig}) for $M_{\rm BH}$; but we do not actually use the
equation until \S\ref{sec:msigma}. 

\subsection{Stellar distribution}
\label{subsec:stdist}

We use the spherical density profile of \citet{hern} to describe
the stars in early-type galaxies at $z=0$. The density in this model
can be written in terms of the total stellar mass, $M_{*,{\rm tot}}$,
and the effective radius, $R_e$:
\begin{align}
\frac{\rho_{*}(r)}{M_{\rm{*,tot}}/R^{3}_{e}} & ~=~  
   \frac{{\mathscr R}^{2}}{2{\pi}}
   \left(\frac{r}{R_{e}}\right)^{-1}
   \left[1 + {\mathscr R}\left(\frac{r}{R_{e}}\right)\right]^{-3}
   ~ ,
\label{eq:srho}
\intertext{where the constant ${\mathscr R}\simeq 1.81527$ (see
  \citealt{hern}). The mass profile,
$\displaystyle M_{*}(r) = \int_{0}^{r}\! 4\pi u^2 \rho_{*}(u)\, du$, is then}
\frac{M_{*}(r)}{M_{\rm{*,tot}}} & ~=~
   \left[\frac{r/R_{e}}{r/R_{e}+1/{\mathscr R}}\right]^{2}.
\label{eq:smass}
\end{align}

Integrating the \citeauthor{hern} $\rho_{*}(r)$ along the line of  
sight gives a surface density profile that closely
approximates the classic $R^{1/4}$ law. Thus, it adequately represents
the typical light distributions in spheroids of mass
$M_{*,{\rm tot}}\sim 10^{10}$--$10^{12}~\msun$, which more generally
follow \citet{sersic}
profiles---$I(R)\sim \exp\left[-(R/R_e)^{1/n}\right]$---with indices
$n\approx3$--7 (e.g., see \citealt{gra97}). These stellar masses
correspond to velocity dispersions
$\sigma_{\rm ap}(R_e)\sim80$--$350~{\rm km~s}^{-1}$
(see Figure \ref{fig:allplots}),
which is the range spanned by the local galaxies that define the black
hole $M$--$\sigma$ relation in Figure~\ref{fig:simple}.

The fine details of the assumed stellar density or mass profile matter
most in our calculations of dimensionless stellar velocity
dispersions $\sigma_{\rm ap}(R_e)\big/\sqrt{GM_{*,{\rm tot}}/R_e}$
using the Jeans equation with model dark matter halos included
(see \S\ref{subsec:apdisp} below). Secondarily, the exact shape of
$\rho_*(r)$ affects the mass ratio $M_{*}(r_{\rm vir})/M_{*}(R_e)$,
which we discuss in \S\ref{subsec:fe}. We examine closely in
\S\ref{sec:modresults} the consequences of using \citeauthor{hern}
profiles for all galaxies in our calculations. In general, it exposes us to
possible errors at the $\sim\!10\%$ level or less.

\subsection{Dark matter distributions}
\label{subsec:dmdist}

Since the dark-matter circular speed $V_{\rm d,pk}$ enters equation
(\ref{eq:msig}) through a high power, it is important that we have a
good idea of how sensitive our results may be to the details of the
dark-matter density profile that we assume. We therefore consider
four different models for spherical halos.
Each of these is a two-parameter model defined by a mass scale and a
radial scale. To treat them uniformly, it is most convenient to
normalise all radii to the point $r_{-2}$ where the logarithmic slope
of the dark-matter density is $d\ln\rho_{\rm d}\big/d\ln r=-2$.
Masses are then normalised to the mass enclosed within $r<r_{-2}$.

First, the usual \citetalias{NFW97} profile \citep{NFW96,NFW97} has
density
\begin{align}
\rho_{\rm d}(r) & ~\propto~ \left(\frac{r}{r_{-2}}\right)^{-1}
   \left(1+\frac{r}{r_{-2}}\right)^{-2} ~ ,
\label{eq:nfwrho}
\intertext{which yields the mass profile}
\frac{M_{\rm d}(r)}{M_{\rm d}(r_{-2})} & ~=~
   \frac{\ln\left(1+r/r_{-2}\right) - (r/r_{-2})(1+r/r_{-2})^{-1}}
        {\ln(2)-1/2} ~ .
\label{eq:nfwmass}
\intertext{The circular-speed curve of the halo {\it alone}, i.e.,
  $V_{\rm d}^2(r)=GM_{\rm d}(r)\big/r$, is then given by}
\frac{V_{\rm d}^2(r)}{V_{\rm d}^2(r_{-2})} & ~=~
   \frac{\ln\left(1+r/r_{-2}\right) - (r/r_{-2})(1+r/r_{-2})^{-1}}
        {(r/r_{-2})\left[\ln(2)-1/2\right]} ~ ,
\label{eq:nfwvc}
\intertext{which peaks at the radius}
\frac{r_{\rm pk}}{r_{-2}} & ~\simeq~ 2.16258 ~ .
\label{eq:nfwrpk}
\end{align}

The second model is that of \citet{hern}, which was first fitted to
simulated dark-matter halos by \citet{DNC}. This has the same central
density cusp ($\rho_{\rm d}\rightarrow r^{-1}$) as an
\citetalias{NFW97} halo, but a
steeper large-radius slope ($\rho_{\rm d}\rightarrow r^{-4}$ rather
than $r^{-3}$) and hence a finite, rather than divergent, total
mass. When written in terms of $r_{-2}$ and $M(r_{-2})$ rather 
than the effective radius and total mass, the model is
\begin{align}
\rho_{\rm d}(r) & ~\propto \left(\frac{r}{r_{-2}}\right)^{-1}
   \left(1+\frac{1}{2}\frac{r}{r_{-2}}\right)^{-3}
\intertext{and}
\frac{M_{\rm d}(r)}{M_{\rm d}(r_{-2})} & ~=~
   9\left(\frac{r/r_{-2}}{2+r/r_{-2}}\right)^2 ~ ,
\label{eq:hernmass}
\intertext{giving a circular-speed curve,}
\frac{V_{\rm d}^2(r)}{V_{\rm d}^2(r_{-2})} & ~=~
   \frac{9\,r/r_{-2}}{\left(2+r/r_{-2}\right)^2} ~ ,
\label{eq:hernvc}
\intertext{with a peak at radius}
\frac{r_{\rm pk}}{r_{-2}} & ~=~ 2 ~ .
\label{eq:hernrpk}
\end{align}

The third model is one from the family developed by
\citet{dm}, which reproduces the universal power-law behaviour of
``pseudo'' phase-space density profiles,
$\rho_{\rm d}(r)\big/\sigma_{\rm d}^3(r)$, in simulated dark-matter
halos. This model fits the resolved parts of the density profiles
alone better
than either the \citetalias{NFW97} or \citeauthor{hern} profiles, and
about as well 
as the \citet{ein65} density profiles with
$\rho_{\rm d}(r)\sim \exp(-r^{\alpha})$, first advocated in this context
by \citet{gra06}. The \citeauthor{dm} density is
\begin{align}
\rho_{\rm d}(r) & ~\propto~ \left(\frac{r}{r_{-2}}\right)^{-7/9}
   \left[1+\frac{11}{13}\left(\frac{r}{r_{-2}}\right)^{4/9}\right]^{-6}
   ~ .
\label{eq:dmrho}
\intertext{This has a slightly shallower central cusp than the
\citetalias{NFW97} or
  \citeauthor{hern} profiles and a large-radius fall-off,
  $\rho_{\rm d}\rightarrow r^{-31/9}$, which is steeper than
  \citetalias{NFW97}
  (resulting in a finite total halo mass) but shallower than
  \citeauthor{hern}. The mass profile is then}
\frac{M_{\rm d}(r)}{M_{\rm d}(r_{-2})} & ~=~
   \left[\frac{24\left(r/r_{-2}\right)^{4/9}}
              {13+11\left(r/r_{-2}\right)^{4/9}}\right]^5
\label{eq:dmmass}
\intertext{and the circular-speed curve is}
\frac{V_{\rm d}^2(r)}{V_{\rm d}^2(r_{-2})} & ~=~
   \left[\frac{24\left(r/r_{-2}\right)^{11/45}}
              {13+11\left(r/r_{-2}\right)^{4/9}}\right]^5 ~ ,
\label{eq:dmvc}
\intertext{which reaches its peak value at}
\frac{r_{\rm pk}}{r_{-2}} & ~=~ \left(\frac{13}{9}\right)^{9/4}
  ~\simeq~ 2.28732 ~ .
\label{eq:dmrpk}
\end{align}

Finally, the halo model of \citet{bur95} has a
constant-density core that appears more suited to the dynamics
of some low-mass galaxies (e.g., \citealt{bns97}), and a large-radius
fall-off that is the same as \citetalias{NFW97}. Here, the density
is
\begin{align}
\rho_{\rm d}(r) & ~\propto~
   \left(1+{\mathscr R}\frac{r}{r_{-2}}\right)^{-1} 
   \left(1+{\mathscr R}^2\frac{r^2}{r_{-2}^2}\right)^{-1} ~ ,
\label{eq:burkrho}
\intertext{with ${\mathscr R}\simeq 1.52138$. The corresponding mass
  profile is}
\frac{M_{\rm d}(r)}{M_{\rm d}(r_{-2})} & ~=~ \notag\\
\intertext{%
\vspace*{-\baselineskip}\begin{equation}
     \hfill
     \frac{\ln\left[\left(1+{\mathscr R} r/r_{-2}\right)
            \sqrt{1+{\mathscr R}^2 \left(r/r_{-2}\right)^2}\,\right]
          - \tan^{-1}\left({\mathscr R} r/r_{-2}\right)}
        {\ln\left[\left(1+{\mathscr R}\right)
            \sqrt{1+{\mathscr R}^2}\,\right] - \tan^{-1}({\mathscr R})}
     ~ ,
\label{eq:burkmass}
\end{equation}
which gives a circular-speed curve,}
\frac{V_{\rm d}^2(r)}{V_{\rm d}^2(r_{-2})} & ~=~ \notag\\
\intertext{
\vspace*{-\baselineskip}\begin{equation}
     \hfill
     \frac{\ln\left[\left(1+{\mathscr R} r/r_{-2}\right)
            \sqrt{1+{\mathscr R}^2 \left(r/r_{-2}\right)^2}\,\right]
          - \tan^{-1}\left({\mathscr R} r/r_{-2}\right)}
        {(r/r_{-2})\,\left\{
            \ln\left[\left(1+{\mathscr R}\right)
            \sqrt{1+{\mathscr R}^2}\,\right] - \tan^{-1}({\mathscr R})
          \right\}}
     ~ ,
\label{eq:burkvc}
\end{equation}
that peaks at  \vspace*{-1.5\baselineskip}}
\\\frac{r_{\rm pk}}{r_{-2}} & ~\simeq~ 2.13433 ~ .
\label{eq:burkrpk}
\end{align}

\begin{figure}
\begin{center}
\includegraphics[width=83mm]{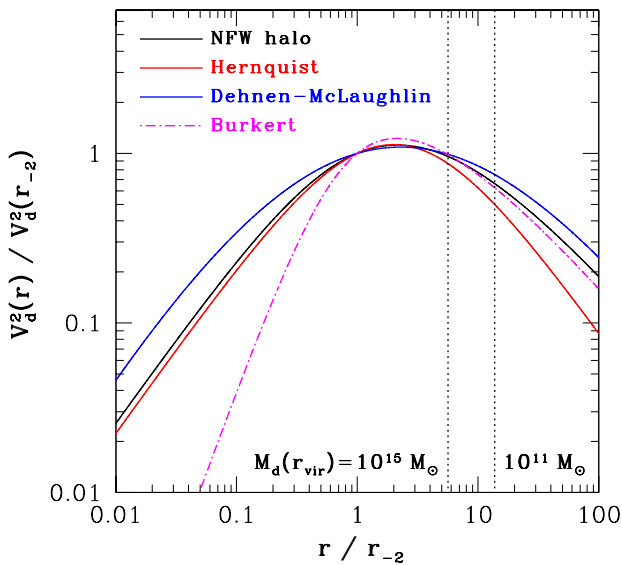}
\end{center}
\caption{Normalised circular-speed curves,
  $V_{\rm d}^2(r)=GM_{\rm d}(r)\big/r$, for the four dark-matter halo
  models we consider. The radius $r_{-2}$ is that where the local
  density slope is $d\ln\rho_{\rm d}\big/d\ln r=-2$.
  The peaks in $V_{\rm d}(r)$ occur at radii near
  $r_{\rm pk}/r_{-2}\approx 2$ in all cases (see text). Broken
  vertical lines show the concentrations $r_{\rm vir}/r_{-2}$ of halos
  with virial masses $M_{\rm d}(r_{\rm vir})=10^{15}~M_\odot$ and 
  $10^{11}~M_\odot$ at $z=0$ (see \S\ref{subsec:conc}). The different
  widths of the circular-speed curves for the different halos lead to
  different values for the baryon fraction inside a
  stellar effective radius (which is typically  in the range
  $R_e/r_{-2}\sim 0.02$--$0.1$; see \S\ref{sec:modresults}), 
  as well as different ratios $V_{\rm d,pk}/\sigma_{\rm ap}(R_e)$.}
\label{fig:vcvpk}
\end{figure}

Figure \ref{fig:vcvpk} shows the circular-speed curves of these
halos, from equations (\ref{eq:nfwvc}), (\ref{eq:hernvc}),
(\ref{eq:dmvc}) and (\ref{eq:burkvc}). Relative to the
\citetalias{NFW97} profile, 
the \citeauthor{hern} curve has a narrower width overall because of its
steeper decline beyond the peak, which follows from its steeper
density profile and convergent mass as
$r\rightarrow \infty$. The \citeauthor{bur95} $V_{\rm d}^2(r)$ profile
is much narrower because of its steeper rise from small $r$,
which is a result of its having a constant-density core rather than a
central density cusp. The \citet{dm} halo has the broadest
circular-speed curve overall, largely because of how slowly its density
profile (which depends on $r^{4/9}$ rather than just $r$) rolls over
from its central cusp with $\rho_{\rm d}(r)\sim r^{-7/9}$  to its
power-law behaviour $\rho_{\rm d}(r)\sim r^{-31/9}$ at large
radii. In the analysis of \S\ref{sec:modresults}, these features
ultimately affect not only the ratio
$V_{\rm d,pk}/\sigma_{\rm ap}(R_e)$, but also the self-consistent 
value of $M_{*}(R_e)\big/M_{\rm d}(R_e)$, the stellar mass fraction
inside the effective radius.

\subsection{Stellar-to-dark matter mass ratios}
\label{subsec:fvir}

The global ratio of stellar to dark-matter mass in galaxies is a
strong and non-monotonic function of halo mass that changes with
redshift. \citet{behr13} compare several derivations of this function
at $z\approx 0$ by different groups using different methods. 
In this paper, we adopt a parametrisation from \citet{mos10}.

\citeauthor{mos10} assign one central galaxy to each virialised halo
(which might be a sub-halo within a larger structure having its own
central galaxy) in $\Lambda$CDM simulations of structure formation
with $\Omega_{m,0}=0.26$, $\Omega_{\Lambda,0}=0.74$ and
$H_0=72~{\rm km~s}^{-1}~{\rm Mpc}^{-1}$.
The stellar mass of any central galaxy is determined by the
virial mass of its parent halo according to a prescription that is
required ultimately to give agreement between the simulations and the
observed galaxy luminosity function. They fit their results,
for the central-galaxy mass fraction $M_*/M_{\rm d}$ within the virial
radius $r_{\rm vir}$ at $z=0$, with a double power-law function:
\begin{align} 
\frac{M_{*}(r_{\rm vir})}{M_{\rm d}(r_{\rm vir})} & ~=~
  0.0564 \left\{
    \left[\frac{M_{\rm d}(r_{\rm vir})}{7.66\times10^{11}~M_\odot}\right]^{-1.06} 
  \right.
  \notag\\
 & \qquad\qquad\qquad \left. ~+~
  \left[\frac{M_{\rm d}(r_{\rm vir})}{7.66\times10^{11}~M_\odot}\right]^{+0.556}
  \right\}^{-1}
\label{eq:mos}
\end{align}
(see their equation [2] and their Table 6).
We discuss the virial radii themselves in the next subsection.
Stellar mass fractions inside any other radius follow
self-consistently from specifications of the stellar and
dark-matter density profiles, as \S\ref{sec:modresults} will detail.

Equation (\ref{eq:mos}) represents an average trend; scatter
around can be expected, for example, as a result of differences in the
merger histories of halos with the same mass at $z=0$.
\citet{mos10} and \citet{behr13}
show that the relation is in good overall agreement with other
theoretical work and/or with data, for halo virial masses 
$10^{11}\,M_\odot\la M_{\rm d}(r_{\rm vir})\la 10^{15}\,M_\odot$. 
This corresponds to stellar masses
$5\times10^8~M_\odot \la M_{*}(r_{\rm vir}) \la 10^{12}~M_\odot$ for the
central galaxies. The brightest galaxies used to define the observed
$M$--$\sigma$ relation are at the upper end of this range.

Equation (\ref{eq:mos}) does not attempt to account for the total
baryonic mass within the virial radius of any 
halo; it is only for stellar mass, and only that concentrated at
the centre. There will be significantly more baryonic mass
in large (cluster-sized) halos especially, in the form of
intracluster light and X-ray gas, and in the stars of
galaxies inside virialised sub-halos. We discuss this
further in \S\ref{sec:modresults} and conclude that the
complication of additional baryons can safely be ignored for our
purposes.

\subsection{Virial radii and cosmological parameters}
\label{subsec:rvir}

We use the fitting formula of
\citet[][see their equation {\mbox{[6]}}]{bnn98} to calculate the
overdensity, relative to the critical density, of a virialised sphere
at redshift $z$ in a flat universe with a
cosmological constant ($\Omega_{m}+\Omega_{\Lambda}=1$):
\begin{align}
\Delta_{\rm vir}(z) & ~\equiv~
   \frac{2\,G M(r_{\rm vir})}{H^2(z)\,r_{\rm vir}^3} 
\notag \\
 & ~\simeq~ 18\pi^2
    \,-\, 82\,\frac{1-\Omega_{m,0}}{\left[H(z)\big/H_0\right]^2}
    \,-\, 39\,\frac{\left(1-\Omega_{m,0}\right)^2}
                   {\left[H(z)\big/H_0\right]^4} ~ ,
\label{eq:deltavir}
\intertext{with}
\left[\frac{H(z)}{H_0}\right]^2 & ~=~
   1 \,+\, \Omega_{m,0}\left[\left(1+z\right)^3 - 1\right] ~ .
\label{eq:hubz}
\end{align}
Rearranging the definition of $\Delta_{\rm vir}$ yields a
convenient relationship between virial radius and virial mass at
arbitrary redshift:
\begin{equation}
\left[\frac{M(r_{\rm vir})}{M_\odot}\right]
\left[\frac{r_{\rm vir}}{{\rm kpc}}\right]^{-3} ~=~
  1166.1\,h_0^2\,\Delta_{\rm vir}(z)\,\left[\frac{H(z)}{H_0}\right]^2
  ~ ,
\label{eq:mrvir}
\end{equation}
where
$h_0\equiv H_0/\left(100~{\rm km~s}^{-1}~{\rm Mpc}^{-1}\right)$
as usual. This form is also useful for calculating $M/r^3$ of spheres
with other overdensities $\Delta$ besides the virial value
[e.g., $\Delta(z)\equiv 200$].

Whenever we use any of equations
(\ref{eq:deltavir})--(\ref{eq:mrvir}), we take cosmological parameters
from the 
{\it Planck} 2013 results \citep{planck14}: $h_0=0.67$ with
$\Omega_{m,0}=0.32$ (which includes a baryon density of
$\Omega_{b,0}=0.049$) and $\Omega_{\Lambda,0}=0.68$. 

\subsection{Halo concentrations}
\label{subsec:conc}

By the concentration of a dark-matter halo, we specifically mean the
ratio of $r_{\rm vir}$ (within which, the mean overdensity is given by
equation [\ref{eq:deltavir}]) to $r_{-2}$ (where the slope of the
density profile is $d\ln\rho_{\rm d}/d\ln r=-2$).
It is also common in the literature to define concentration as the
ratio of $r_{200}$ (within which, the mean overdensity is
$\Delta= 200$) to $r_{-2}$.
Either way, $N$-body simulations of CDM structure formation
consistently show that, at least for low redshifts, more massive halos
have lower concentrations on average. We need to
take account of this in order to infer the location
and the value of the maximum circular
speed in any dark-matter halo with a given virial radius and mass.

\citet{dnm14} give a fitting formula for the concentrations
$r_{\rm vir}/r_{-2}$ of simulated halos with masses
$10^{11}\,M_\odot \la M_{\rm d}(r_{\rm vir}) \la 10^{15}\,M_\odot$ at
redshifts $0\le z \le 5$ in a {\it Planck} cosmology. Namely,
\begin{align}
\log\left[\frac{r_{\rm vir}}{r_{-2}}\right] & ~\simeq~
      a \,-\,  b \log
      \left[ \frac{M_{\rm d}(r_{\rm vir})}{10^{12}\,h_0^{-1}~M_\odot}
      \right]
\label{eq:dut}
\intertext{with}
 a & ~=~ 0.537 \,+\, 0.488\exp\left(-0.718\,z^{1.08}\right)
\notag \\
 b & ~=~ 0.097 \,-\, 0.024\,z ~ .
\notag
\end{align}
Again, we set $h_0=0.67$ whenever we use this equation.
Simulated halos scatter around the average trend
at the level of a few tens of percent in $r_{\rm vir}/r_{-2}$ for a
fixed virial mass and redshift (\citealt{bul01,dnm14}). 

\citeauthor{dnm14} obtain equation (\ref{eq:dut})
by fitting \citetalias{NFW97} density profiles to their simulated halos in
order to measure the radius $r_{-2}$. They also investigate the use of
\citet{ein65} profiles instead (which are more like the \citealt{dm}
halos that we explore) to fit for $r_{-2}$ in estimating the
alternative concentration $r_{200}/r_{-2}$.
Their results suggest that concentration values depend on the
choice of model for the dark-matter density profile, but only at the
$\la\!10\%$ level for halos with $M_{\rm d}(r_{\rm vir})\ga
10^{12}~M_\odot$ at $z=0$. We apply equation 
(\ref{eq:dut}) in our models regardless of what model we assume for
$\rho_{\rm d}(r)$ and simply accept that there is a modest
uncertainty associated with doing so.

The two vertical lines in Figure \ref{fig:vcvpk} show the
concentrations according to equation (\ref{eq:dut}) for halos with
virial masses at $z=0$ of $M_{\rm d}(r_{\rm vir})=10^{11}\,M_\odot$
(having $r_{\rm vir}/r_{-2}=13.8$) and
$10^{15}\,M_\odot$ (having $r_{\rm vir}/r_{-2}=5.64$).
Equation (\ref{eq:mos}) gives the corresponding stellar masses of the
central galaxies as $M_{*}(r_{\rm vir})=6.3\times10^8\,M_\odot$ and 
$1.0\times10^{12}\,M_\odot$. This emphasises the degree to which
$V_{\rm d,pk}$---the key predictor of self-limited SMBH masses in the
simple feedback model behind equation (\ref{eq:msig})---reflects
conditions far outside the stellar distributions of normal galaxies
(generally, $R_e/r_{-2}\sim 0.02$--$0.1$; see
\S\ref{sec:modresults}). 

Equation (\ref{eq:dut}) has been derived from simulations of
strictly baryon-free halos. This is not an issue for
our modelling, precisely because the equation describes halos
on large scales $r > r_{-2} \gg R_e$, well away from any
regions that might have been altered significantly by the presence of
stars.

\subsection{Halo progenitors}
\label{subsec:haloev}

If the central black hole in a protogalaxy ended its main,
quasar phase of accretion growth at a redshift $z>0$, with a mass
$M_{\rm BH}$ determined by the circular speed $V_{\rm d,pk}$ in the
dark-matter halo at that time, then we need to relate that earlier
$V_{\rm d,pk}$ to the value at $z=0$ (in order ultimately
to link it and $M_{\rm BH}$ to a stellar velocity dispersion at $z=0$).

From $N$-body simulations and merger trees of $\Lambda$CDM halos with
virial masses at $z=0$ in the range
$10^{11}\,M_\odot \la M_{\rm d,vir}(0) \la 10^{15}\,M_\odot$,
\citet{vdb14} extract for each halo the redshift $z_{1/2}$ at which
its {\it most massive progenitor} had a virial mass
$M_{\rm d,vir}(z_{1/2})=0.5\,M_{\rm d,vir}(0)$. Given the bottom-up
nature of structure formation in CDM cosmologies, $z_{1/2}$ is a
decreasing function of $M_{\rm d,vir}(0)$ in general. We 
have fitted the median dependence shown in Figure 4 of
\citeauthor{vdb14} with the function 
\begin{equation}
z_{1/2} ~=~
   2.05\left[
      \frac{M_{\rm d,vir}(0)}{10^{12}\,h_0^{-1}~M_\odot}
   \right]^{-0.055} - ~ 1 ~ ,
\label{eq:zhalf}
\end{equation}
again taking $h_0=0.67$ from the {\it Planck} cosmology.
Once again, there is intrinsic scatter around this overall trend.

Given $z_{1/2}$, we then approximate the virial mass of the most
massive progenitor of a halo at any other redshift by the exponential
function \citep[see also, e.g.,][]{zha09},
\begin{equation}
\frac{M_{\rm d,vir}(z)}{M_{\rm d,vir}(0)} ~=~
\exp{\left[-\,\frac{\ln(2)}{z_{1/2}}~z\right]} ~ .
\label{eq:mdvirz}
\end{equation}
Equations (\ref{eq:zhalf}) and (\ref{eq:mdvirz}) together give curves
of $M_{\rm d,vir}(z)\big/M_{\rm d,vir}(0)$ versus
$M_{\rm d,vir}(0)$ that, for redshifts $z\la 5$, compare well to
the curves plotted by \citet{vdb14} directly from the simulations they
analyse (e.g., see their Figure 2).
  
\begin{figure}
\begin{center}
\includegraphics[width=83mm]{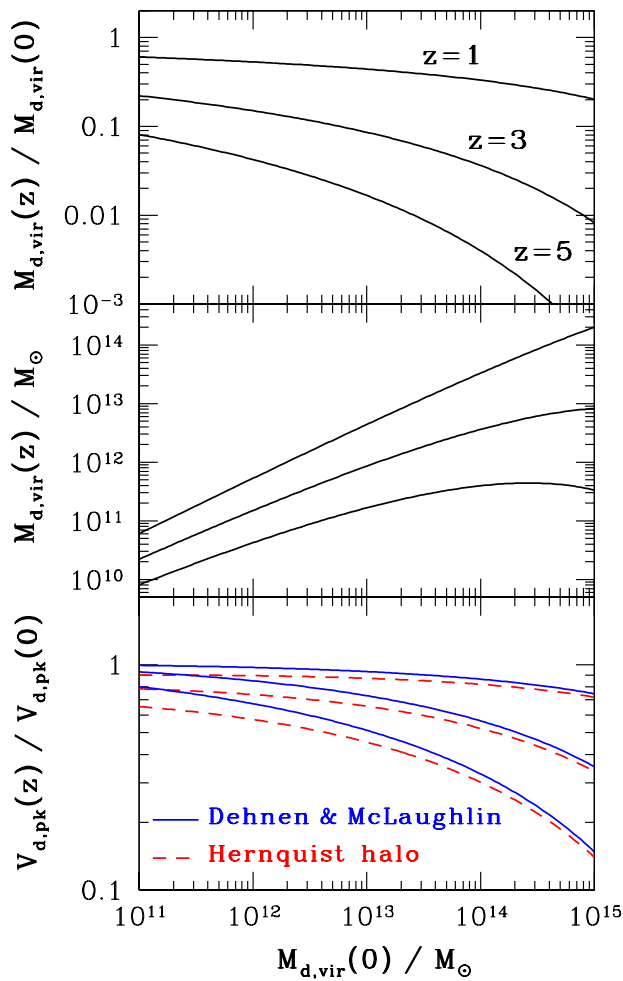}
\end{center}
\caption{{\it Top panel}: Relative virial masses
  $M_{\rm{d,vir}}(z)/M_{\rm{dvir}}(0)$ for the most massive
  progenitors of halos with masses $M_{\rm{d,vir}}(0)$ at
  $z=0$, as given by equations (\ref{eq:zhalf}) and
  (\ref{eq:mdvirz}). From top to bottom, the curves are for the
  progenitors at fixed redshifts $z=1,3$ and 5. 
  {\it Middle panel}: Virial masses of the most
  massive progenitor halos at $z=1,\,3$ and 5 (for the curves
  from top to bottom) plotted directly against the $z=0$ halo mass.
  {\it Bottom panel}: Peak circular speeds $V_{\rm{d,pk}}(z)$ in the
  most massive progenitors at $z=1,\,3$ and 5, relative to the peak
  speeds $V_{\rm{d,pk}}(0)$ in the final halos at $z=0$, from equation
  (\ref{eq:vpkz}). The solid (blue) lines  
  are for halos with a \citet{dm} density profile and the dashed (red)
  lines are for halos with a \citet{hern} profile. These bracket the
  corresponding curves for \citetalias{NFW97} and \citet{bur95} halos
  at the same redshifts.}
\label{fig:vpkz}
\end{figure}

To obtain the evolution of the peak circular speed in the
most massive progenitor of a halo, we first write (for any $z$)
\begin{equation}
\frac{V_{\rm d,pk}^2}{V_{\rm d,vir}^2} ~\equiv~
   \frac{V_{\rm d}^2(r_{\rm pk})}{V_{\rm d}^2(r_{\rm vir})} ~=~
   \frac{g(r_{\rm pk}/r_{-2})}{g(r_{\rm vir}/r_{-2})}
\label{eq:vpkza}
\end{equation}
where $g(r/r_{-2})$ is one of the normalised circular-speed
curves shown in Figure \ref{fig:vcvpk} and written on the 
right-hand sides of equations (\ref{eq:nfwvc}), (\ref{eq:hernvc}),
(\ref{eq:dmvc}) and (\ref{eq:burkvc}) above. Then, since the ratio
$r_{\rm pk}/r_{-2}$ is independent of redshift (it is fixed by
assuming a basic form for the dark-matter density profile),
we have
\begin{align}
\frac{V_{\rm d,pk}^2(z)}{V_{\rm d,pk}^2(0)} & ~=~ 
   \frac{g\left[\left(r_{\rm vir}/r_{-2}\right)_{z=0}\right]}
        {g\left[\left(r_{\rm vir}/r_{-2}\right)_{z}\right]} ~\times~
   \frac{V_{\rm d,vir}^2(z)}{V_{\rm d,vir}^2(0)}
\notag \\
  & ~=~
    \frac{g\left[\left(r_{\rm vir}/r_{-2}\right)_{z=0}\right]}
         {g\left[\left(r_{\rm vir}/r_{-2}\right)_{z}\right]} ~\times~ 
    \notag \\
  & \qquad\qquad
    \left[\frac{M_{\rm{d,vir}}(z)}{M_{\rm{d,vir}}(0)}\right]^{2/3}
    \left[\frac{\Delta_{\rm{vir}}(z)}{\Delta_{\rm vir}(0)}\right]^{1/3}
    \left[\frac{H(z)}{H_0}\right]^{2/3}
    ~ ,
\label{eq:vpkz}
\end{align}
where the last line uses the fact that
$V_{\rm d}^2(r)\propto M_{\rm d}(r)\big/r$ and brings in equation
(\ref{eq:mrvir}). For any choice of dark-matter halo model, and
thus of the function $g(r/r_{-2})$, the right-hand side of equation
(\ref{eq:vpkz}) is known in terms of $z$ and $M_{\rm d,vir}(0)$, via
equations (\ref{eq:zhalf}) and (\ref{eq:mdvirz}) plus equations
(\ref{eq:deltavir}), (\ref{eq:hubz}) and  (\ref{eq:dut}).

The upper panel of Figure \ref{fig:vpkz} shows the virial masses at
$z=1,\,3$ and 5, relative to the $z=0$ virial masses, for the
most massive progenitors of halos spanning the range of
$M_{\rm d,vir}(0)$ investigated by \citet{vdb14}. 
The middle panel shows the masses of the largest progenitors at
$z=1,\,3$ and 5 directly as functions of the halo mass at $z=0$. The
curves in these plots 
are the same for any model of the halo density profile.
The lower panel of Figure \ref{fig:vpkz} shows the ratio of
progenitor-to-present $V_{\rm d,pk}$ at $z=1,\,3$ and 5 against
the $z=0$ virial mass, calculated using equation (\ref{eq:vpkz}).
These curves depend on the halo density profile. For clarity, we
only show results assuming either 
a \citet{dm} or a \citet{hern} density profile, so $g(r/r_{-2})$ is
given either by equation (\ref{eq:dmvc}) or by equation
(\ref{eq:hernvc}).

It is worth noting here the gradual flattening
towards higher masses of the curves for $M_{\rm d,vir}(z)$ versus
$M_{\rm d,vir}(0)$ in the middle panel of Figure \ref{fig:vpkz}, and
how the flattening sets in at more modest halo
masses for larger $z$. This is a generic feature of structure formation
by hierarchical merging. Halos in any given mass range at $z=0$
have progenitors drawn from increasingly narrow mass
ranges, on average, at increasingly high redshift; and this
narrowing is more pronounced as a function of $z$ for higher-mass
halos, because more of their growth has occurred more recently.

Precise numbers---such as the possible value of
a maximum mass for the largest progenitors suggested by the $z=5$
curve in Figure \ref{fig:vpkz}---are specific to the dependence  
of $z_{1/2}$ on $M_{\rm d,vir}(0)$ in our equation (\ref{eq:zhalf}). 
That and equation (\ref{eq:mdvirz}) only give an {\it approximation}
to the numerical results of \citet{vdb14} for the {\it median}
most-massive progenitors of halos with
$10^{11}\,M_\odot\la M_{\rm d,vir}(0)\la 10^{15}\,M_\odot$.
Fine details following from them are not definitive, especially at
the highest end of the $z=0$ mass range. However, the flattening of
$M_{\rm d,vir}(z)$ as a function of $M_{\rm d,vir}(0)$ is
qualitatively robust. It ultimately has some implications for the
shape of the black hole $M$--$\sigma$ relation at high
$\sigma$-values, which we discuss further in \S\ref{sec:msigma}. 

In the bottom panel of Figure \ref{fig:vpkz}, at any fixed redshift
the different halo models give greater differences in
$V_{\rm d,pk}(z)/V_{\rm d,pk}(0)$ for lower virial masses.
This is because lower-mass halos generally have higher concentrations
$r_{\rm vir}/r_{-2}$, and therefore higher ratios of
$r_{\rm vir}/r_{\rm pk}$ (see equation [\ref{eq:dut}]). Thus, the
ratio $V_{\rm d,pk}/V_{\rm d,vir}$ is more sensitive in lower-mass
halos to the model-dependent steepness of the circular-speed curve at
radii $r>r_{\rm pk}$. But
$V_{\rm d,vir}^2\propto M_{\rm d,vir}(z)\big/r_{\rm vir}(z)$ is
independent of the halo density profile, and so only 
$V_{\rm d,pk}$ is actually model-dependent. Since
\citetalias{NFW97} and \citet{bur95} halos have circular-speed curves
that are intermediate in steepness to \citeauthor{dm} and 
\citeauthor{hern} models beyond $r_{\rm pk}$ (see Figure
\ref{fig:vcvpk}), the curves for $V_{\rm d,pk}(z)/V_{\rm d,pk}(0)$
versus $M_{\rm d,vir}(0)$ in these other models lie between the two
shown in Figure \ref{fig:vpkz}. 

\section{Galaxy and halo scalings at \,{\boldmath$\MakeLowercase{z}=0$}}
\label{sec:modresults}

A two-component model for a spherical galaxy is formally defined by
four parameters:  
$R_{e}$ and $M_{*,{\rm tot}}$ for the stars, which we assume here
to follow \citet{hern} density profiles (summarised in
\S\ref{subsec:stdist}), plus $r_{-2}$ and $M_{\rm d}(r_{-2})$
for a dark-matter profile
(described in \S\ref{subsec:dmdist}). However, there are
interdependences between these parameters: $R_e$ and
$M_{*,{\rm tot}}$ are correlated (discussed just below),
while the radii and masses of dark-matter halos are connected to
each other and to $M_{*,{\rm tot}}$ by cosmological simulations (the
stellar mass fractions in \S\ref{subsec:fvir} and the concentrations
in \S\ref{subsec:conc}). These dependences allow the models to
be put in terms of a single independent parameter, which we choose to
be $M_{*,{\rm tot}}$.  

Figure \ref{fig:allplots} shows the average trends for various galaxy 
properties versus $M_{*,{\rm tot}}$ at $z=0$,
together in some cases with data from the literature. In this Section
we detail the procedures leading to these plots.
In \S\ref{sec:msigma}, we fold in the
redshift evolution of $V_{\rm d,pk}$ (from \S\ref{subsec:haloev})
to apply equation (\ref{eq:msig}) for predicted black hole
masses and consider the empirical correlation between $M_{\rm BH}$
and the stellar $\sigma_{\rm ap}(R_e)$.

Our goal here is to establish representative trend-line
relationships between various stellar and halo properties. Scatter
around the trends is inevitable, and it can contain physical
information, but in this paper we set aside the task of
characterising or explaining any scatter in detail.

\subsection{Stellar masses and effective radii}
\label{subsec:re}

\begin{figure*}
\begin{center}
\includegraphics[width=\textwidth]{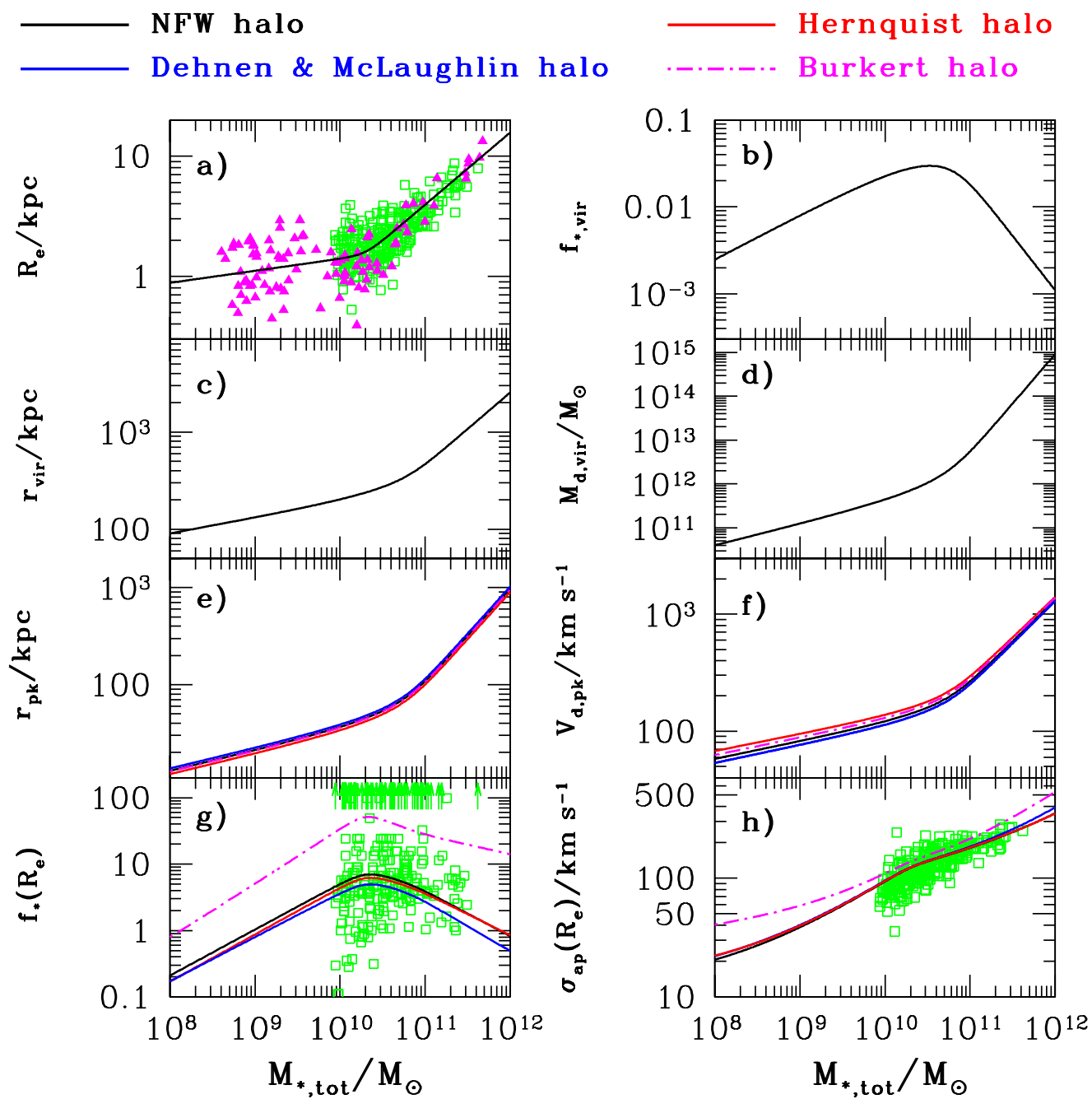} 
\end{center}
\caption{Model scaling relations for stellar and dark-matter halo
  properties versus total stellar mass, $M_{*,{\rm tot}}$, in
  spherical galaxies at $z=0$. With the exception of the curve in
  panel (a), the low-mass extensions of these models to
  $M_{*,{\rm tot}} \la 5\times10^9\,M_\odot$ (stellar velocity
  dispersions $\sigma_{\rm ap}(R_e)\la 60~{\rm km~s}^{-1}$) should be
  viewed with some caution, as discussed in \S\ref{subsec:discussz0}.
  \newline
  {\it Panel (a)}: Stellar effective radius, $R_e$. Data points
  represent galaxies in the ATLAS$^{\rm 3D}$ survey
  (\citealt{capI}; green squares)
  and the {\it ACS} Virgo Cluster Survey
  (\citealt{chen}; magenta triangles). See
  \S\ref{subsec:re} for details. 
  {\it Panel (b)}: Ratio $f_{*,{\rm vir}}$ of stellar-to-dark matter
  mass within the virial radius; see \S\ref{subsec:fvir} and
  \S\ref{subsec:virial}.
  {\it Panel (c)}: Virial radius, $r_{\rm vir}$; see
  \S\ref{subsec:virial}. 
  {\it Panel (d)}: Mass of dark matter within the
  virial radius, $M_{\rm d,vir}$; see \S\ref{subsec:virial}. 
  {\it Panel (e)}: Radius $r_{\rm pk}$ where the dark-matter
  circular-speed curve peaks. The different coloured curves are for
  four different models of the dark-matter density profile. See
  \S\ref{subsec:dmdist} and \S\ref{subsec:dmvcpk} for details.
  {\it Panel (f)}: Peak value of the dark-matter circular speed,
  $V_{\rm d,pk}$, assuming each of the four different dark-matter halo
  models; see \S\ref{subsec:dmvcpk}.
  {\it Panel (g)}: Ratio $f_*(R_e)$ of stellar mass to dark matter
  mass within a sphere of radius $r<R_e$, for each of the four
  different halo models; see \S\ref{subsec:fe}. Data points are from
  dynamical modelling by the ATLAS$^{\rm 3D}$ survey
  \citep{capxv,capxx}; arrows at the top of the panel represent
  galaxies consistent in their analysis with having no dark matter
  inside $R_e$.
  {\it Panel (h)}: Stellar velocity dispersion $\sigma_{\rm ap}(R_e)$
  within an aperture of radius $R_e$. Data points are taken from the
  ATLAS$^{\rm 3D}$ survey. See \S\ref{subsec:apdisp} for details.}
\label{fig:allplots}
\end{figure*}

Panel (a) of Figure \ref{fig:allplots} plots effective radius against
total stellar mass for local early-type galaxies in two datasets:
258 systems from the ATLAS$^{\rm 3D}$ survey
(squares: \citealt{capI,capxv,capxx}) and
100 from the {\it ACS} Virgo Cluster Survey
(ACSVCS, triangles: \citealt{cote04,chen}).

In each case, the effective radii are tabulated by the original
authors, either in kpc directly or as angular sizes along with
the distances to individual galaxies.
To estimate the stellar masses, we have taken integrated luminosities
provided by the authors and calculated mass-to-light ratios
using the single-burst population-synthesis models of
\cite{Mar98,Mar05} assuming stellar ages of 9 Gyr and a
\cite{kroupa01} stellar initial mass function (IMF). The masses in
these $M/L$ ratios include both luminous stars and dark
remnants. We have also used \citet{bc03} models to confirm that
extended star formation lasting as long as 6~Gyr gives
the same $M/L$ values, to within $\la5\%$, when the mean stellar age is
9~Gyr.

\citet{capI} give $K$-band absolute magnitudes for
galaxies in the ATLAS$^{\rm 3D}$ survey.
At an age of 9~Gyr and for metallicities
$-1.7\le {\mbox{[Z/H]}} \le +0.3$, the mass-to-light ratios tabulated by
\citet{Mar05} are
$0.93\ga M_*/L_K\ga 0.82~M_\odot\,L_\odot^{-1}$. We therefore
adopt a constant $M_*/L_K\equiv 0.88~M_\odot\,L_\odot^{-1}$ for all of
the ATLAS$^{\rm 3D}$ galaxies. This value changes by approximately
$\pm15\%$ if the mean age of the stars is changed by
$\pm2$~Gyr.

\citet{chen} give $g$-band apparent magnitudes and $(g-z)$
colours for the ACSVCS galaxies. Combining these with
surface-brightness fluctuation distances from \citet{blake09}
allows us to calculate absolute $z$-band magnitudes. Then, for
metallicities $-1.7\le {\mbox{[Z/H]}} \le +0.3$, a
\citeauthor{kroupa01} IMF and an age of 9 Gyr, the
\citeauthor{Mar05} models give
$1.40 \la M_*/L_z \la 2.0~M_\odot\,L_\odot^{-1}$.
We have used a single $M_*/L_z\simeq 1.7~M_\odot\,L_\odot^{-1}$
for all of the ACSVCS galaxies to plot the points in panel (a)
of Figure \ref{fig:allplots}. Again, this changes by $\pm15\%$--20\%
if the assumed age is  changed by $\pm2$~Gyr.

The line going through the $R_{e}$~vs~$M_{*,{\rm tot}}$ data in Figure
\ref{fig:allplots} is a parametrisation of the average correlation,
\begin{equation}
\frac{R_{e}}{{\rm kpc}} ~=~ 
 1.5~
 \left(\frac{M_{\rm{*,tot}}}{2{\times}10^{10}~M_\odot}\right)^{0.1}
 \left[
 1+\left(\frac{M_{\rm{*,tot}}}{2{\times}10^{10}~M_\odot}\right)^{5}
 \right]^{0.1} ~~ ,
\label{eq:rems}
\end{equation}
which we decided by eye. Roughly equal numbers of
ATLAS$^{\rm 3D}$\,+\,ACSVCS data points lie above and below this line.
A $\pm 20\%$ change in adopted mass-to-light ratios (whether due
to a different assumed mean age or a different star formation history)
results in a $\pm 20\%$ change to the mass
scale in equation (\ref{eq:rems}).

The ATLAS$^{\rm 3D}$ sample covers the full range of stellar
masses, $10^{10}\,M_\odot \la M_{*,{\rm tot}} \la 10^{12}\,M_\odot$,
of the local galaxies that define the black hole
$M$--$\sigma$ relation. As mentioned in \S\ref{subsec:stdist}, the 
light profiles in this mass range can generally be
fitted by \citet{sersic} models with indices $n\approx 3$--7, all of
which can be approximated adequately, for our purposes, by a
\citet{hern} profile in projection. The ACSVCS galaxies include many
with $M_{*,{\rm tot}} < 10^{10}~M_\odot$, where surface-brightness
profiles are increasingly better fitted by lower-index
\citeauthor{sersic}  
functions tending towards exponentials. We have included these
systems mainly to ensure that our analysis
incorporates the change in slope that they show in the
$R_e$--$M_{*,{\rm tot}}$ correlation. In all of what follows, we
address with some care the extent to which our results might (or may
not) be put in error by assuming \citeauthor{hern} stellar-density
profiles for all systems.

\subsection{Virial radii and halo virial masses}
\label{subsec:virial}

For any value of $M_{*,{\rm tot}}$, equation (\ref{eq:rems}) gives a
typical value for $R_e$. Assuming a \citeauthor{hern} density
profile for the stars we can then write, for the ratio of
stellar-to-dark matter mass within the virial radius of a galaxy,
\begin{align}
f_{*,{\rm vir}} & ~\equiv~
   \frac{M_{*}(r_{\rm vir})}{M_{\rm d}(r_{\rm vir})} ~=~
       \frac{M_{*,{\rm tot}}}{M_{\rm d,vir}}
       \left[\frac{r_{\rm vir}/R_e}{r_{\rm vir}/R_e + 1/\mathscr{R}}
             \right]^2
\label{eq:fvir1}
\intertext{with $\mathscr{R}\simeq 1.81527$ (see equation
  [\ref{eq:smass}]). Understanding the dark-matter mass to be that of
  the main halo {\it centred} on the stars in the galaxy,
  $f_{*,{\rm vir}}$ is additionally constrained by
  cosmological simulations, as discussed in
  \S\ref{subsec:fvir} and represented by equation (\ref{eq:mos}) above
  from \citet{mos10}. Repeating this for convenience, at $z=0$ we
  have}
f_{*,{\rm vir}} & ~=~ 
  0.0564 \left\{
    \left[\frac{M_{\rm d, vir}}{7.66\times10^{11}~M_\odot}\right]^{-1.06} 
  \right.
  \notag\\
 & \qquad\qquad\qquad \left. ~+~
  \left[\frac{M_{\rm d, vir}}{7.66\times10^{11}~M_\odot}\right]^{+0.556}
  \right\}^{-1} ~ .
\label{eq:fvir2}
\intertext{Finally, if the total mass within $r_{\rm vir}$ is simply
  the sum of the dark matter plus the stars in the central galaxy,
  i.e., $M(r_{\rm vir})=M_{\rm d,vir}\left(1+f_{*,{\rm vir}}\right)$,
  then the definition of $r_{\rm vir}$ in equation (\ref{eq:mrvir})
  gives (at $z=0$ for the 2013 {\it Planck} cosmological parameters)}
f_{*,{\rm vir}} & ~=~ 0.0544
   \left[\frac{r_{\rm vir}}{100~{\rm kpc}}\right]^3
   \left[\frac{M_{\rm d,vir}}{10^{12}~M_\odot}\right]^{-1}
   ~-~ 1 ~ .
\label{eq:fvir3}
\end{align}
Solving equations (\ref{eq:fvir1})--(\ref{eq:fvir3}) for all of
$f_{*,{\rm vir}}$, $r_{\rm vir}$ and $M_{\rm d,vir}$ as functions of
$M_{*,{\rm tot}}$ gives the curves shown in panels (b), (c) and (d) of
Figure \ref{fig:allplots}. These are independent of any assumptions
about the internal density profiles of the halos.

The peak in $f_{*,{\rm vir}}$ in panel (b), at a value of
$\simeq\!0.03$ for $M_{*,{\rm tot}}\simeq 3.4\times10^{10}~M_\odot$
or $M_{\rm d,vir} \simeq 1.1\times10^{12}~M_\odot$, comes directly
from the form of equation (\ref{eq:fvir2}) taken from
\citet{mos10}. It is intriguing that the mass scale of
this peak is close to the mass where the empirical 
$R_e$--$M_{*,{\rm tot}}$ correlation changes slope (equation
[\ref{eq:rems}]), but we do not pursue this issue here.
The immediate point is that $f_{*,{\rm vir}}$ decreases rapidly towards 
higher masses, such that the halos around central galaxies with
$M_{*,{\rm tot}} \ga 10^{11}~M_\odot$ have
$M_{\rm d,vir}\ga 10^{13}~M_\odot$ and $r_{\rm vir}\ga 500~{\rm kpc}$.
They encompass entire groups and clusters.

For the massive systems in particular, there may be baryons that
reside in the halos but are not
associated directly with the stars of the central
galaxy---intracluster light and gas, and the stars in any off-centre
satellite galaxies. Equation (\ref{eq:fvir3}) for the virial radius 
takes no account of any such ``extra'' baryons. To do so properly
would require additionally constraining the global baryon fraction in
galaxy clusters, which is itself a mass-dependent
quantity (see, e.g., \citealt{giodini09,mcgaugh10,zhang11}). However,
in no case would the total virial mass be increased by more than
$\simeq\!15\%$ (this being the cosmic average baryon fraction,
$\Omega_{b,0}/\Omega_{m,0}$), and hence the virial radius would not
increase by more than $\simeq\!5\%$. We therefore ignore the
complication as far as $r_{\rm vir}$ is concerned.

Then, over the range of galaxy masses shown in Figure
\ref{fig:allplots}, we find that $110 \la r_{\rm vir}/R_e \la 170$.
As a result, the stellar mass inside the virial radius is
$M_{*}(r_{\rm vir}) \ga 0.99\,M_{*,{\rm tot}}$ in all cases, and
equation (\ref{eq:fvir1}) says that 
$f_{*,{\rm vir}}\simeq M_{*,{\rm tot}}/M_{\rm d,vir}$ with only a very
weak dependence on $r_{\rm vir}/R_e$. The mass of dark
matter {\it alone} within $r_{\rm vir}$ is then determined 
(through equation [\ref{eq:fvir2}]) by
$M_{*,{\rm tot}}$ almost 
independently of $r_{\rm vir}$. Thus, our values for $M_{\rm d,vir}$
would not be changed discernibly by having additional baryons
distributed in the halos outside of the central galaxies.

These conclusions still hold if the stars in the central galaxies 
are described by \citeauthor{sersic} models that depart significantly
from \citeauthor{hern} profiles in projection, so long as 
$M_*(r)$ still essentially converges within $r\la 100\,R_e$. Hence,
the curves for $f_{*,{\rm vir}}$, $r_{\rm vir}$
and $M_{\rm d,vir}$ versus $M_{*,{\rm tot}}$ in Figure
\ref{fig:allplots} are insensitive to the choice of stellar density
profile.

\subsection{Peak halo circular speeds}
\label{subsec:dmvcpk}

With virial radii and dark-matter virial masses known as functions of
$M_{*,{\rm tot}}$, the scale $r_{-2}$ follows from equation
(\ref{eq:dut}) in \S\ref{subsec:conc} for the concentration
$r_{\rm v ir}/r_{-2}$ versus 
$M_{\rm d,vir}$ \citep{dnm14}, evaluated at $z=0$. The location
of the peak of the dark-matter circular-speed curve then
comes from the ratio $r_{\rm pk}/r_{-2}$ specific to a choice of
$\rho_{\rm d}(r)$ for the dark matter (one of equations
[\ref{eq:nfwrpk}], [\ref{eq:hernrpk}], [\ref{eq:dmrpk}] or
[\ref{eq:burkrpk}] in \S\ref{subsec:dmdist}). Panel (e) of Figure
\ref{fig:allplots} shows the final curves of $r_{\rm pk}$ versus
$M_{*,{\rm tot}}$ for all four of the halo profiles we are
considering. There is little difference between the curves because we
have assumed the same $(r_{\rm vir}/r_{-2})$ versus $M_{\rm d,vir}$
relation for all halo models, and because
$r_{\rm pk}/r_{-2} = 2$--2.3 for all of
them. They are also essentially independent of the form of the stellar
density profile, because the underlying curves of $r_{\rm vir}$ and
$M_{\rm d,vir}$ versus $M_{*,{\rm tot}}$ are. Ultimately, we have
approximately $15\la r_{\rm pk}/R_{e}\la 70$ and
$0.14\la r_{\rm pk}/r_{\rm vir} \la 0.40$ for stellar masses in the
range $10^8\,M_\odot\la M_{*,{\rm tot}}\la 10^{12}\,M_\odot$.

The peak value of the dark-matter circular speed is obtained as
\begin{equation}
V_{\rm d, pk}^2 ~=~
   \frac{\,V_{\rm d}^2(r_{\rm pk})\big/V_{\rm d}^2(r_{-2})\,}
        {\,V_{\rm d}^2(r_{\rm vir})\big/V_{\rm d}^2(r_{-2})\,}
   ~ \frac{G\,M_{\rm d,vir}}{r_{\rm vir}} ~ .
\label{eq:vcpk}
\end{equation}
The normalised circular-speed profiles
$V_{\rm d}^2(r)\big/V_{\rm d}^2(r_{-2})$ 
for different halo models are shown in Figure \ref{fig:vcvpk}
and given in equations (\ref{eq:nfwvc}), (\ref{eq:hernvc}),
(\ref{eq:dmvc}) and (\ref{eq:burkvc}) of \S\ref{subsec:dmdist}.
Evaluating the appropriate one of these at $r_{\rm pk}/r_{-2}$ and
$r_{\rm vir}/r_{-2}$ after choosing a density profile 
$\rho_{\rm d}(r)$, and then
folding in the dependences of $M_{\rm d,vir}$ and $r_{\rm vir}$ on
$M_{*,{\rm tot}}$, yields $V_{\rm d, pk}$ at any given total stellar
mass. The results are shown in panel (f) of Figure
\ref{fig:allplots}.

The curves for $V_{\rm d,pk}$ versus $M_{*,{\rm tot}}$ are again
insensitive to the use of a \citeauthor{hern} profile for the 
stellar distributions. The differences between them come from the
(small) differences in the values of $r_{\rm pk}/r_{-2}$ in the
different halo models, and the (larger) differences in the widths of
the normalised circular-speed curves between $r_{\rm pk}$ and
$r_{\rm vir}$, as seen in Figure \ref{fig:vcvpk}.
The differences are greater for systems with smaller
$M_{*,{\rm tot}}$ because those halos are less massive and have
higher concentrations on average, with larger ratios
$r_{\rm vir}/r_{\rm pk}$ and hence ratios
$V_{\rm d}(r_{\rm pk})/V_{\rm d}(r_{\rm vir})$ that are more sensitive
to the shape of the circular-speed curve at large radii in a halo.

It is clear that the circular speeds $V_{\rm d,pk}$ for the most
massive model galaxies, which represent those defining the 
upper end of the observed black hole $M$--$\sigma$ relation, will far
exceed the stellar velocity dispersions measured within $R_e$ in the
real systems. This is because the dark-matter halos
centred on such massive galaxies correspond to entire clusters.
It is also why the naive substitution
$V_{\rm d,pk}=\sqrt{2}\,\sigma_{\rm ap}(R_e)$, inspired by the
singular isothermal sphere, cannot suffice for a proper comparison of
a prediction like equation (\ref{eq:msig}) to the
$M$--$\sigma$ data (cf.~Figure \ref{fig:simple}). At the same time,
the most massive 
halos are the ones that will have grown the most at low redshifts,
{\it after} the epoch of peak quasar activity that may have mainly
determined self-regulated black hole masses.
Hence it is essential that $V_{\rm d,pk}$ be calculated in the
{\it progenitors} of halos if equation (\ref{eq:msig}) is to be
assessed self-consistently.

\subsection{Stellar mass fractions inside {\boldmath$R_e$}}
\label{subsec:fe}

The ratio of stellar mass to dark-matter mass contained within radius
$r$ in a galaxy with a specified total stellar mass can be written as
\begin{equation}
f_{*}(r) ~\equiv~ \frac{M_*(r)}{M_{\rm d}(r)} ~=~
   f_{*,{\rm vir}}\,
   \frac{\,M_*(r)\big/M_*(r_{\rm vir})\,}
        {\,M_{\rm d}(r)\big/M_{\rm d}(r_{\rm vir})\,} ~ .
\label{eq:fofr}
\end{equation}
Here, $f_{*,{\rm vir}}$ is known from above as a function of
$M_{*,{\rm tot}}$. The normalised stellar mass profile
$M_*(r)/M_*(r_{\rm vir})$ comes from equation (\ref{eq:smass}) for a
\citeauthor{hern} density profile and is determined by $M_{*,{\rm tot}}$
because $R_e$ and $r_{\rm vir}$ are. Once a dark-matter halo model has
been chosen, the mass profile $M_{\rm d}(r)/M_{\rm d}(r_{\rm vir})$
follows from one of equations (\ref{eq:nfwmass}), (\ref{eq:hernmass}),
(\ref{eq:dmmass}) or (\ref{eq:burkmass}) and is also determined
by $M_{*,{\rm tot}}$ because that fixes the concentration
$r_{\rm vir}/r_{-2}$.

The function $f_*(r)$ enters into the Jeans equation for calculations
of the stellar velocity dispersion in \S\ref{subsec:apdisp}. First,
however, we evaluate it specifically at the radius $r=R_e$ for
galaxies with a range of stellar masses, in order to compare our results
with some additional data.

\cite{capxv,capxx} have used dynamical (Jeans) modelling to estimate
the ratio of dark-to-total mass within a {\it sphere} of radius
$r=R_{e}$ for each of the ATLAS$^{\rm 3D}$ galaxies. This fraction,
which they denote $f_{\rm dm}$, is related to our stellar-to-dark mass
ratio within $r<R_e$ by $f_*(R_e) = f_{\rm dm}^{-1} - 1$. Although
the \citeauthor{capxv} modelling assumes that dark-matter halos
have \citetalias{NFW97} density profiles, their results are not
sensitive to this detail, since usually
$M_{\rm d}(R_e) < M_*(R_e)$ by  factors of several in their
galaxies---see \citet{capxv} for further details. 

Panel (g) of Figure \ref{fig:allplots} shows the $f_{*}(R_e)$ data for
258 ATLAS$^{\rm 3D}$ galaxies (arrows at the top of the panel indicate
galaxies for which the modelling by \citeauthor{capxv} is consistent
with no dark matter inside $r<R_e$). The curves show the
typical $f_*(R_e)$ expected at a given $M_{*,{\rm tot}}$ on the basis
of our equation (\ref{eq:fofr}), for each of the four different
dark-matter halo profiles. 

These curves depend on the stellar density profile as
$f_*(R_e)\propto M_*(R_e)/M_*(r_{\rm vir}) \simeq M_*(R_e)/M_{*,{\rm tot}}$.
In the mass range
$M_{*,{\rm tot}}\ga 10^{10}\,M_\odot$, describing the stars by
\citeauthor{sersic} models with $3\la n\la 7$ rather than by
\citeauthor{hern} models alters $M_*(R_e)/M_{*,{\rm tot}}$, and
hence $f_*(R_e)$, by less than 5\%. Much lower-mass galaxies, which
have no $f_*(R_e)$ data in Figure 
\ref{fig:allplots} and are not represented in the empirical
$M$--$\sigma$ relation, will have closer to exponential
surface-brightness profiles. For these, $M_*(R_e)/M_{*,{\rm tot}}$ and
$f_*(R_e)$ are lower than the \citeauthor{hern} model values, but by
no more than $\simeq\!20\%$.

The curves are rather more sensitive to the choice of dark-matter halo
profile, in particular to how steeply the enclosed mass
$M_{\rm d}(r)$ decreases inwards to $r\rightarrow 0$. This is
reflected in the shapes of the circular-speed curves in Figure
\ref{fig:vcvpk}. For a given value of
$M_{*,{\rm tot}}$, and hence $M_{\rm d}(r_{\rm vir})$,
\citetalias{NFW97} and  
\citeauthor{hern} halos have similar values for
$M_{\rm d}(R_e)/M_{\rm d}(r_{\rm vir})$, and thus for $f_{*}(R_e)$,
because of their identical central structures.
\citet{dm} halos have higher 
$M_{\rm d}(R_e)/M_{\rm d}(r_{\rm vir})$ and lower
$f_*(R_e)$ for the same stellar mass, because they have significantly
shallower mass profiles than either
\citetalias{NFW97} or \citeauthor{hern} halos.
The much steeper $M_{\rm d}(r)$ or
$V_{\rm d}^2(r)$ profiles in the constant-density cores
of \citet{bur95} models put substantially
more dark matter at large radii in these halos, giving lower values
of $M_{\rm d}(R_e)/M_{\rm d}(r_{\rm vir})$ and higher $f_*(R_e)$ for
a fixed $M_{*,{\rm tot}}$.

The three dark-matter halos with central density cusps all imply
$f_*(R_e)$ values that are broadly consistent with the data
in Figure \ref{fig:allplots}(g)
for systems with $M_{*,{\rm tot}}\ga 10^{10}\,M_\odot$. However, the
cored halo of \citet{bur95} is incompatible with these data.
This is a valuable check on our calculations, and an argument for not
considering \citeauthor{bur95} halos further in the context of the
black hole $M$--$\sigma$ relation for intermediate- and high-mass
galaxies. But it is not surprising, since the
\citeauthor{bur95} model was originally proposed only in connection
with dwarf spheroidal galaxies, not regular ellipticals.

\subsection{Stellar velocity dispersions}
\label{subsec:apdisp}

To calculate stellar velocity dispersions, we solve the isotropic
Jeans equation including contributions to the gravitational potential
from the dark matter, the stars {\it and} the accumulated
ejecta from stellar winds and supernovae over the lifetime
of a galaxy. Assuming that these ejecta are confined to the
central regions of the overall potential well in relatively large
galaxies, we approximate their mass profile as
$M_{\rm ej}(r) \approx F_{\rm ej}M_*(r)$ with $F_{\rm ej}$ a
constant. The value of $F_{\rm ej}$ comes from the same
single-burst population-synthesis models that we used in
\S\ref{subsec:re} to calculate stellar mass-to-light ratios. Namely,
for a \citet{kroupa01} stellar IMF and stellar ages greater than
several Gyr, \citet{Mar05} gives the ratio of current-to-initial
mass in stars (and remnants) as $\simeq\!0.58$. Thus, in our notation,
$\left(1+F_{\rm ej}\right)\simeq 1/0.58$.
The value of $F_{\rm{ej}}$ is 
robust to any changes in the star formation history, with a $<2\%$
increase for extended star formation.

With dimensionless radii, stellar densities and one-dimensional
velocity dispersions defined as 
\begin{equation}
\widetilde{r}~\equiv~\frac{r}{R_e}
~;\qquad
\widetilde{\rho}_*~\equiv~
   \frac{\rho_*}{M_{*,{\rm tot}}/R_e^3}
~;\qquad
\widetilde{\sigma}_*^2~\equiv~
       \frac{\sigma_*^2}{GM_{*,{\rm tot}}/R_e}
\notag
\end{equation}
the isotropic and spherical Jeans equation is 
\begin{equation}
\frac{d}{d\widetilde{r}}
  \Big[\widetilde{\rho}_*(\widetilde{r})~
       \widetilde{\sigma}_*^2(\widetilde{r})
  \Big] \,=\, -\, \frac{\widetilde{\rho}_*(\widetilde{r})}
                     {\widetilde{r}^2}\,
                \frac{M_*(\widetilde{r})}{M_{*,{\rm tot}}}\,
                \left[ \left(1+F_{\rm ej}\right) \,+\,
                       \frac{1}{f_*(\widetilde{r})} \right]  ~ .
\label{eq:jeans}
\end{equation}
The profiles $\widetilde{\rho}_*(\widetilde{r})$ and
$M_*(\widetilde{r})/M_{*,{\rm tot}}$ are given by equations
(\ref{eq:srho}) and (\ref{eq:smass}) in \S\ref{subsec:stdist} for a
\citeauthor{hern} model, while $(1+F_{\rm ej})=1/0.58$ as just
mentioned. The function
$f_*(\widetilde{r}) \equiv M_*(\widetilde{r})
   \big/M_{\rm d}(\widetilde{r})$
is known in full for any specific value of $M_{*,{\rm tot}}$ (and
choice of dark-matter density profile) as discussed in
\S\ref{subsec:fe}.
Subject to the boundary condition that
$\widetilde{\rho}_*\,\widetilde{\sigma}_*^2 \rightarrow 0$ as
$\widetilde{r}\rightarrow \infty$, equation (\ref{eq:jeans}) can
therefore be solved for the dimensionless
$\sigma_*^2\big/\left(GM_{*,{\rm tot}}/R_e\right)$ as a function
of $r/R_e$ in a galaxy with any given total stellar mass.

The aperture velocity dispersion over a circular disc on the
plane of the sky comes from projecting $\sigma_*^2(r)$
along the line of sight and then taking a luminosity-weighted
average. Defining the dimensionless projected radius
$\widetilde{R}\equiv R/R_e$, the stellar surface-density profile is
first obtained as
\begin{equation}
\widetilde{\Sigma}_*(\widetilde{R}) ~\equiv~
   \frac{\Sigma_*(R)}{M_{*,{\rm tot}}/R_e^2} ~=~
  2 ~ \int_{\widetilde{R}}^{\infty} \widetilde{\rho}_*(\widetilde{r})~
      \frac{\widetilde{r}~d\widetilde{r}}
           {(\widetilde{r}^2 - \widetilde{R}^2)^{1/2}} ~ ;
\label{eq:surf}
\end{equation}
then the projected stellar velocity-dispersion profile is
\begin{equation}
\widetilde{\sigma}_{\rm p}^2(\widetilde{R}) ~=~
   \frac{2}{\widetilde{\Sigma}_*(\widetilde{R})}~
   \int_{\widetilde{R}}^{\infty}
       \widetilde{\rho}_*(\widetilde{r})~
       \widetilde{\sigma}_*^2(\widetilde{r})~
       \frac{\widetilde{r}~d\widetilde{r}}
            {(\widetilde{r}^2 - \widetilde{R}^2)^{1/2}} ~ ;
\label{eq:siglos}
\end{equation}
and the aperture dispersion within projected radius
$R_{\rm ap}$ is
\begin{align}
\frac{\sigma_{\rm ap}^2(R_{\rm ap})}{GM_{*,{\rm tot}}/R_e} & ~=~
  \left[\, \int_{0}^{R_{\rm ap}/R_e}
           \widetilde{\sigma}_{\rm p}^2(\widetilde{R})~
           \widetilde{\Sigma}_*(\widetilde{R})~
           \widetilde{R}\,d\widetilde{R} \,\right]
\notag \\
 & \null \qquad\qquad~~\times~~
   \left[\, \int_{0}^{R_{\rm ap}/R_e}
           \widetilde{\Sigma}_*(\widetilde{R})~
           \widetilde{R}\,d\widetilde{R} \,\right]^{-1} ~ .
\label{eq:sigap}
\end{align}
The right-hand side of this is determined entirely by
$M_{*,{\rm tot}}$ once a halo model has been chosen and a value of
$R_{\rm ap}$ specified. Setting $R_{\rm ap}=R_e$ yields the model
$\sigma_{\rm ap}$ that corresponds to the measured velocity
dispersions in the \citet{mcconnellma} compilation of SMBH
$M$--$\sigma$ data.

Panel (h) of Figure \ref{fig:allplots} shows the calculated
$\sigma_{\rm ap}(R_e)$ versus $M_{*,{\rm tot}}$ for each of the four
different dark-matter halo models. The points are data for the
ATLAS$^{\rm 3D}$ galaxies, taken again from
\cite{capI,capxv,capxx} (the ACSVCS 
galaxies included in the plot of $R_e$ versus $M_{*,{\rm tot}}$ do not
have published velocity dispersions). All of the cuspy halos give
curves that run through the middle of the $\sigma_{\rm ap}(R_e)$
data, while the cored \citet{bur95} halo predicts velocity dispersions
that are higher for a given $M_{*,{\rm tot}}$. A \citeauthor{bur95}
halo has relatively more of its mass at larger radii than the cuspy
halos do. The unprojected $\sigma_*(r)$ is substantially higher
around and beyond $r\sim R_e$ as a result, which inflates the
line-of-sight dispersion even inside $R_e$ and boosts the aperture
dispersion noticeably.

The dimensionless aperture dispersion inside $R_e$ for a
self-consistent \citeauthor{hern} sphere of stars only, with no ejecta
or dark matter ($F_{\rm ej}=0$ and $1/f_*(r)\equiv 0$), is
$\sigma_{\rm ap}(R_e)/\left(GM_{*,{\rm tot}}/R_e\right)^{1/2}
     \simeq 0.389$. 
Based on this, the dispersion with ejecta and dark matter included can
be usefully approximated by the function
\begin{equation}
\frac{\sigma_{\rm ap}(R_e)}{\sqrt{GM_{*,{\rm tot}}/R_e}}
   ~\approx~ 0.389\,
   \sqrt{\,\left(1+F_{\rm ej}\right) + \frac{0.86}{f_*(R_e)}\,} ~~~ ,
\label{eq:sigapprox}
\end{equation}
where the term under the square-root represents the ratio of an
``effective'' total mass to the total stellar mass.
This formula reproduces the values from our full calculations with 
relative error $<\!2.5\%$ for any $f_*(R_e)>0.1$ in any of an NFW,
\citeauthor{hern} or \citeauthor{dm} halo.

We have also calculated
$\sigma_{\rm ap}(R_e)/(GM_{*,{\rm tot}}/R_e)^{1/2}$ for self-gravitating
\citet{sersic} $R^{1/n}$ spheres without any
dark matter. For indices $n\la 5$---which apply to
giant ellipticals and dwarfs with masses down to
$M_{*,{\rm tot}}\sim 10^8$--$10^9~M_\odot$---we find 
$0.36\la \widetilde{\sigma}_{\rm ap}(R_e) \la 0.43$, as compared to
$\widetilde{\sigma}_{\rm ap}(R_e)\simeq 0.389$ for the
\citeauthor{hern} model. Thus, over most of the mass range in Figure
\ref{fig:allplots}, the model curves for $\sigma_{\rm ap}(R_e)$ 
are vulnerable at only the $\la\!10\%$ level to bias (a slight tilt)
resulting from our use of a \citeauthor{hern} profile to describe all
of the stellar distributions. Very massive ellipticals with
$M_{*,{\rm tot}} \ga 2$--$3\times10^{11}\,M_\odot$ are generally fitted
by \citeauthor{sersic} indices $n\approx5$--7, for which
$\sigma_{\rm ap}(R_e)/(GM_{*,{\rm tot}}/R_e)^{1/2}\simeq 0.43$--0.49
rather than 0.389. However, a small compensation in our
parametrisation of $R_e$ versus $M_{*,{\rm tot}}$ at high masses then
suffices to yield essentially the same $\sigma_{\rm ap}(R_e)$ as the
curve in Figure \ref{fig:allplots}(h).

\subsection{Discussion}
\label{subsec:discussz0}

\subsubsection{Dwarf galaxies}
\label{subsubsec:dwarf}

There are more 
physical considerations than the validity of a \citeauthor{hern}
profile for the stellar distribution, which affect how well
our models might be able to describe galaxies with
stellar masses less than a few $\times10^9\,M_\odot$.

In order to calculate velocity dispersions in
\S\ref{subsec:apdisp}, we 
assumed that stellar ejecta are retained at the bottom
of any galaxy's potential well. However, supernova-driven winds will
have expelled the ejecta from many dwarf ellipticals to far beyond the
stellar distributions. In this case, $F_{\rm ej}=0$ in equations
(\ref{eq:jeans}) and (\ref{eq:sigapprox}) is more appropriate than
$(1+F_{\rm ej})=1/0.58$. This lowers the expected
$\sigma_{\rm ap}(R_e)$ by $\approx\!30\%$ at a given $M_{*,{\rm tot}}$ for
a given halo density profile.

On the other hand, the same galactic winds may cause changes in the
central structures of the dark-matter halos of dwarfs, from initially
steep density cusps to shallower profiles perhaps closer to the
\citet{bur95} model (e.g., \citealt{bns97,pont12}); while subsequent
tidal stripping could have led to further modifications
at large radii in the halos. Substantial, systematic alterations to
the dark-matter density profiles may impact the values we infer for
$V_{\rm d,pk}$, $f_*(R_e)$ and $\sigma_{\rm ap}(R_e)$
from a given $M_{*,{\rm tot}}$, $R_e$ and $M_{\rm d,vir}$.
And in any case, the relationship connecting $M_{*,{\rm tot}}$ to
$M_{\rm d,vir}$ in equation (\ref{eq:mos}), from \citet{mos10},
may itself be in error if extrapolated to halo masses much below
$M_{\rm d,vir}\la 10^{11}~M_\odot$ (see \citealt{behr13}).

All in all, while the model curves in Figure
\ref{fig:allplots} can be viewed as broadly indicative of the
situation for dwarf galaxies, they should also be seen as provisional
in that regime.
More comprehensive modelling is required to be confident of how these
kinds of average trends extrapolate to stellar masses much less 
than several $\times10^{9}\,M_\odot$ (or, roughly,
$\sigma_{\rm ap}(R_e)\la 60$--$70~{\rm km~s}^{-1}$).

\subsubsection{Intracluster baryons}
\label{subsubsec:icm}

As already discussed in \S\ref{subsec:virial}, we can safely ignore
any small differences that intracluster
baryons (whether gas or stars) might make to the virial radii and
masses we calculate for halos centred on the most massive galaxies.
Equation (\ref{eq:sigapprox}) in \S\ref{subsec:apdisp} now provides a
way to assess the effects of intracluster
baryons on the stellar velocity dispersions in the central
galaxies of groups and clusters.

If additional baryonic mass is distributed spatially like the dark
matter, then it can be accounted for in the Jeans equation
(\ref{eq:jeans}), and hence in equation (\ref{eq:sigapprox}),
by decreasing $f_*(r)\equiv M_*(r)/M_{\rm d}(r)$ by a constant factor.
This factor will be largest if the global baryon fraction in
a halo is equal to the cosmic average value 
but only a trace amount is actually contained in
the central galaxy itself. Thus, an ``effective'' $f_*(r)$ in the
Jeans equation might
be lower than the \citet{mos10} value by a factor of
$\left(1-\Omega_{b,0}/\Omega_{m,0}\right)^{-1}$ at most, which is
$\simeq\!1.18$ for a 2013 {\it Planck} cosmology. This could
plausibly be the situation in halos with
$M_{\rm d}(r_{\rm vir})\sim10^{15}~M_\odot$ (which have
$M_{*,{\rm tot}}\sim 10^{12}~M_\odot$ for the central galaxy),
but the total baryon fraction decreases systematically with
decreasing (sub-)halo mass
\citep[e.g.,][]{gonzalez13,zhang11,mcgaugh10}. In galaxy-sized halos, it
is generally consistent with the mass of stars,
remnants and stellar ejecta in the galaxy proper, which we
have already accounted for fully.

The maximum effect on $\sigma_{\rm ap}(R_e)$ in the central galaxy can
be estimated by comparing the value of 
equation (\ref{eq:sigapprox}) with $(1+F_{\rm ej})=1/0.58$ and
$f_*(R_e)=0.5$---the lowest value in any of our models at
$M_{*,{\rm tot}}=10^{12}~M_\odot$ or $M_{\rm d,vir}\simeq
10^{15}~M_\odot$ in Figure \ref{fig:allplots}---to the value using
$f_*(R_e)=0.5/1.18$ instead. The result is an increase of $<\!5\%$ in
the velocity dispersion. This is of the same order as 
the maximum effect on our values for the halo virial radii. 
We have chosen to ignore intracluster baryons altogether rather
than introduce detailed additional modelling just to make adjustments
that are {\it at most} so small.

\subsubsection{Comparisons to individual systems}
\label{subsubsec:real}

In an Appendix, we make some checks on the average scalings
represented in Figure \ref{fig:allplots}, by comparing various numbers
extracted from them to
relevant data in the literature for the Milky Way, M87 and M49 (the
central galaxies of Virgo sub-clusters A and B) and NGC\,4889 (the
brightest galaxy in the Coma Cluster). The stellar masses and velocity
dispersions of these systems span the range covered by the local
early-type galaxies used to define empirical black hole
$M$--$\sigma$ relations. It is notable in particular that, starting
with just the galaxies' total stellar masses, the scalings imply
detailed properties of the {\it cluster-sized} dark-matter halos
around each of M87, M49 and NGC\,4889, which are in reasonably good
agreement with literature values.

\section{The black hole {\boldmath$M$--$\sigma$} relation}
\label{sec:msigma}

The scalings in \S\ref{sec:modresults} give typical virial masses
and peak circular speeds for dark-matter halos, along with
stellar velocity dispersions inside an effective radius, as one-to-one
functions of galaxy stellar mass at $z=0$. Therefore, they can be
re-cast to give $M_{\rm d,vir}(0)$ and $V_{\rm d,pk}(0)$ directly as
functions of the observable $\sigma_{\rm ap}(R_e)$. If a theory ties
$M_{\rm BH}$ to the properties of halos at some time in the past, then in
order to predict the dependence of $M_{\rm BH}$ on
$\sigma_{\rm ap}(R_e)$ (or any other galaxy properties) now, it
is necessary first to connect the halo properties at $z>0$ to those at
$z=0$.

The SMBH--halo relation we examine here is that given by
equation (\ref{eq:msig}) above, from \citet{mcq12}. To repeat,
\begin{equation}
M_{\rm BH} ~\simeq~ 1.14\times10^{8}\,M_\odot\,
         \left(\frac{f_0}{0.2}\right)
         \left(\frac{V_{\rm{d,pk}}}{200~{\rm km s^{-1}}}\right)^{4}
    ~ .
\tag{1}
\label{eq:msigagain}
\end{equation}
As discussed in \S\ref{subsec:mbhvhalo}, this equation is
limited by simplifying assumptions: for example, about the nature of
quasar-mode SMBH feedback (taken to be purely momentum-conserving) 
and the distribution of gas in protogalaxies (taken to be virialised,
with ongoing cosmic infall ignored). Within these limitations it has
the advantage of generality, being applicable to dark matter halos
with any density profile.

In equation (\ref{eq:msigagain}), $V_{\rm d,pk}$ measures
the potential well of a protogalaxy that just fails to contain the
quasar-mode feedback of an SMBH with mass $M_{\rm BH}$. It thus refers
to conditions at a redshift marking the {\it end} of rapid 
SMBH growth by accretion at Eddington
or supercritical rates in a series of gas-rich mergers.
We denote this redshift by $z_{\rm qso}$. It will be different for
different systems, but we expect the general range to coincide with
the epoch of peak quasar number and SMBH accretion-rate
densities in the Universe: namely, $z_{\rm qso}\sim 2{\mbox{--}}4$ in
most cases (e.g., \citealt{rich06,hop07a,del14}; also
\citealt{dim08,sijacki07,sijacki15}).

In this Section, we apply our calculations from \S\ref{subsec:haloev}
to find {\it typical} values of $M_{\rm d,vir}(z_{\rm qso})$ and
$V_{\rm d,pk}(z_{\rm qso})$ for the {\it most massive progenitors} of
halos, and hence estimate an expected $M_{\rm BH}$ in their central
galaxies, as functions of the stellar $\sigma_{\rm ap}(R_e)$ at
$z=0$. This involves an assumption that the most massive progenitor
halo at $z_{\rm qso}>0$ is the one that ultimately defines the centre
of the larger potential well at $z=0$. This is statistically
accurate but not always true in every individual case---see, for
example, the discussion in \citet{vdb14} of the
distinction between ``most massive'' and ``most contributing''
progenitors. Glossing over this subtlety could lead to a small amount
of scatter in the SMBH $M$--$\sigma$ data relative to our final curves.

The model $M_{\rm BH}$--$\sigma_{\rm ap}(R_e)$ relations we obtain do
not include any growth of the SMBH itself at redshifts
$z<z_{\rm qso}$, which can occur by coalescences in gas-poor galaxy
mergers at the centre of a halo. However, this is distinct from the
growth of the halo as a whole; many sub-halos can be accreted at
low redshift that do not sink to the bottom of the potential well
and thus do not grow the central SMBH.
We discuss this further in \S\ref{subsec:msigma}

\subsection{Halo masses and peak circular speeds at {\boldmath$z>0$}}
\label{subsec:mdvirz}

\begin{figure}
\begin{center}
\includegraphics[width=83mm]{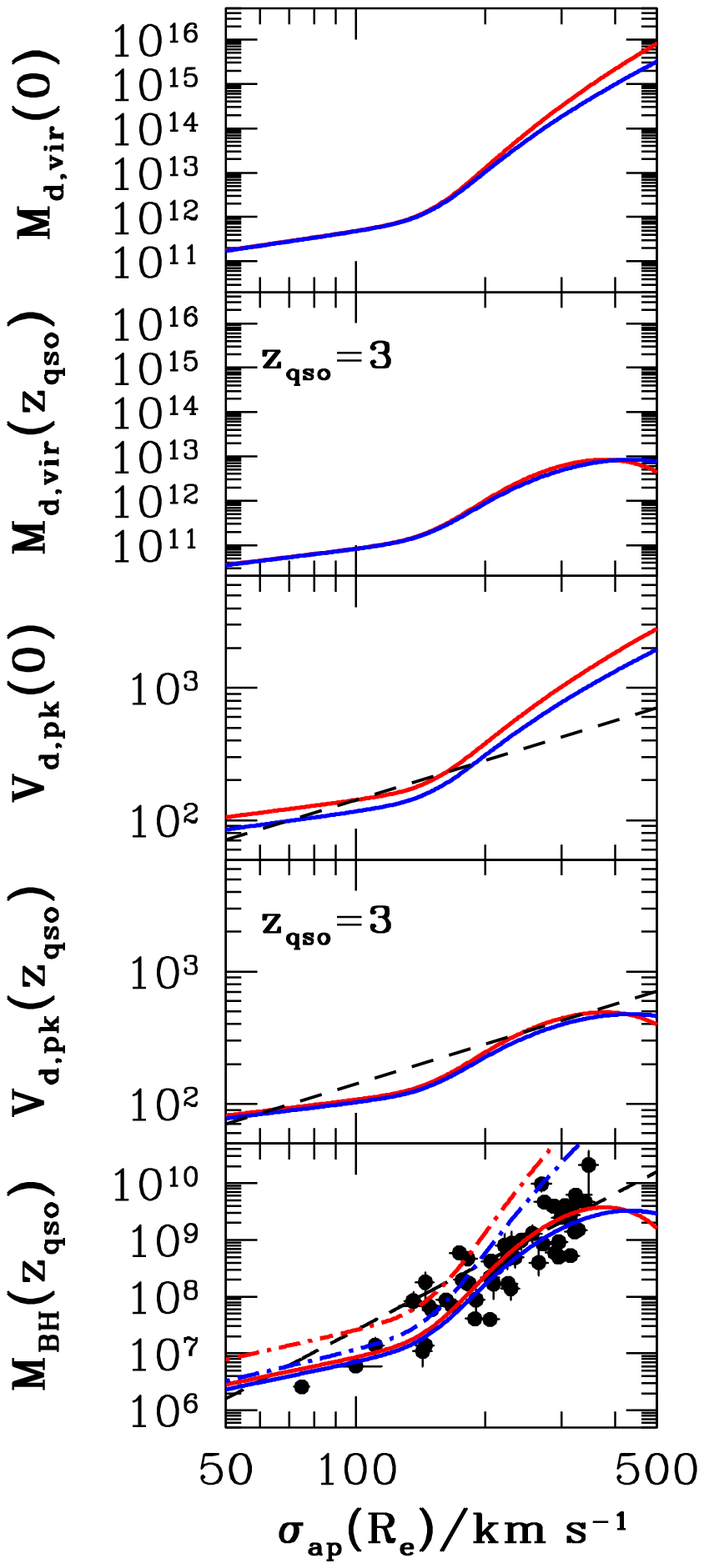}
\end{center}
\caption{{\it Top two panels}: Dark matter virial mass (in $M_\odot$)
  at $z=0$ and at $z_{\rm qso}=3$, versus stellar velocity dispersion
  $\sigma_{\rm ap}(R_e)$ at $z=0$. Blue curves are for galaxy models
  with \citet{dm} halos; red curves have \citet{hern} halos.
  {\it Next two panels}: Peak dark-matter circular speed (in
  km~s$^{-1}$) at $z=0$ and at $z_{\rm qso}=3$, versus
  $\sigma_{\rm ap}(R_e)$ at $z=0$. Blue and red curves correspond
  again to \citeauthor{dm} and \citeauthor{hern} halo density
  profiles. The straight, dashed line shows
  $V_{\rm d,pk}=\sqrt{2}\,\sigma_{\rm ap}(R_e)$.
  {\it Bottom panel}: SMBH mass (in $M_\odot$) calculated from
  equation (\ref{eq:msigagain}) with $f_0=0.18$ using the dark-matter
  $V_{\rm d,pk}$ at $z=0$ (dot-dash blue and red curves) and at
  $z_{\rm qso}=3$ (solid blue and red curves), all plotted against
  $\sigma_{\rm ap}(R_e)$ at $z=0$. The dashed straight line
  is equation (\ref{eq:msigagain}) with
  $V_{\rm d,pk}\equiv\sqrt{2}\,\sigma_{\rm ap}(R_e)$. Data points are
  for the 53 ellipticals and lenticulars in \citet{mcconnellma}.}
\label{fig:mvirvpk}
\end{figure}

The top panel of Figure \ref{fig:mvirvpk} shows the scaling of halo
virial mass at $z=0$ versus stellar velocity dispersion
$\sigma_{\rm ap}(R_e)$ in the central galaxy at $z=0$, obtained
directly from the results of \S\ref{sec:modresults}
[combining panels (d) and (h) of Figure \ref{fig:allplots}].
The next panel down shows $M_{\rm d,vir}$ for
{\it the most massive progenitor of a halo} at redshift
$z_{\rm qso}=3$ [obtained from $M_{\rm d,vir}(0)$ as described in 
\S\ref{subsec:haloev}; see Figure \ref{fig:vpkz}]
against $\sigma_{\rm ap}(R_e)$ in the central galaxy at $z=0$.

The blue curves in Figure \ref{fig:mvirvpk} correspond to \citet{dm}
models for the halo density 
profiles; the red curves, to \citet{hern} models.
These bracket the scalings obtained using \citetalias{NFW97} halo
profiles, while (as discussed in \S\ref{sec:modresults}), the cored
halo profiles of \citet{bur95} are not appropriate in the galaxy mass
range plotted here.
Velocity dispersions $\sigma_{\rm ap}(R_e)\ge 70~{\rm km~s}^{-1}$
at $z=0$ correspond to stellar masses
$M_{*,{\rm tot}}\ga 8$--$9\times10^9\,M_\odot$ at $z=0$.

The next panel in the Figure shows the peak dark-matter circular
speed at $z=0$ versus stellar velocity dispersion at $z=0$, again from
\S\ref{sec:modresults} [combining panels (f) and (h) of Figure
  \ref{fig:allplots}]. Just below this is the scaling of
$V_{\rm d,pk}$ in the most massive progenitor at $z_{\rm qso}=3$
[obtained from $V_{\rm d,pk}(0)$ and $M_{\rm d,vir}(0)$ as in
  \S\ref{subsec:haloev} and Figure \ref{fig:vpkz}]
versus $\sigma_{\rm ap}(R_e)$ in the central galaxy at
$z=0$. The straight, dashed (black) line in these panels
traces out $V_{\rm d,pk}=\sqrt{2}\,\sigma_{\rm ap}(R_e)$. This is
clearly a poor substitute for the actual relationship between the two
velocities at $z=0$ in galaxies with
$\sigma_{\rm ap}(R_e)\ga 200~{\rm km~s}^{-1}$ (or 
$M_{*,{\rm tot}}\ga 3\times10^{11}\,M_\odot$). It does come
closer in this mass range to correctly estimating the dependence of
$V_{\rm d,pk}$ at $z_{\rm qso}=3$ on $\sigma_{\rm ap}(R_e)$ at
$z=0$; but this appears to be entirely coincidental, and the situation
is reversed for $\sigma_{\rm ap}(R_e)\la 200~{\rm km~s}^{-1}$.

At a given value for $\sigma_{\rm ap}(R_e)$, the downwards
``corrections'' to $M_{\rm d,vir}$ and $V_{\rm d,pk}$, from their
values at $z=0$ to the progenitors at $z_{\rm qso}=3$, are
systematically larger for larger systems. This is a restatement of the
flattening towards higher masses in the dependence of
$M_{\rm d,vir}(z)$ on $M_{\rm d,vir}(0)$, which we showed in
Figure \ref{fig:vpkz} and discussed there. Again, it is fundamentally
because in a ($\Lambda$)CDM cosmology, more massive halos were
assembled and virialised more recently. A given range of halo mass or
circular speed at $z=0$ thus corresponds to a narrower range at any
$z_{\rm qso}>0$, and the contrast is greater for higher masses.
In Figure \ref{fig:mvirvpk}, this works to make the 
slopes of $M_{\rm d,vir}$ and $V_{\rm d,pk}$ versus $z=0$
velocity dispersions significantly shallower for the halo progenitors at
$z_{\rm qso}=3$ than for the halos themselves at $z=0$.

The equations from \S\ref{subsec:haloev} that underpin these results
are approximations to the mass accretion histories of simulated halos
in \citet{vdb14}. Those simulations extend up to halo masses
$M_{\rm d,vir}(0)\la 10^{15}\,M_\odot$, corresponding to
stellar $\sigma_{\rm ap}(R_e)\la 350$--$400~{\rm km~s}^{-1}$ at $z=0$.
Beyond this, our analysis is not only approximate but an
extrapolation. Thus, for example, the peaks around
$\sigma_{\rm ap}(R_e)\approx 400~{\rm km~s}^{-1}$ in the panels of
Figure \ref{fig:mvirvpk} for $M_{\rm d,vir}$ and $V_{\rm d,pk}$ at
$z_{\rm qso}=3$ may not be accurate. What is secure is the simple fact
of the relative flatness in these curves for high stellar velocity
dispersions. The same effect must appear to a greater or lesser
degree for any other $z_{\rm qso}>0$, and it directly impacts any
prediction for an observable SMBH $M$--$\sigma$ relation at $z=0$ 
from a model like our equation (\ref{eq:msigagain}) or similar.

\subsection{{\boldmath$M_{\rm BH}$} versus
 {\boldmath$\sigma_{\rm ap}(R_e)$}}
\label{subsec:msigma}

The bottom panel of Figure \ref{fig:mvirvpk} shows SMBH mass versus
$\sigma_{\rm ap}(R_e)$ at $z=0$. The data points are for the
E and S0 galaxies in the compilation of \citet{mcconnellma}.
(Their data for the bulges of late-type galaxies can be seen in Figure
\ref{fig:simple}. We do not show them here because our calculations
for $\sigma_{\rm ap}(R_e)$ versus $M_{*,{\rm tot}}$ do not
allow for discs.)
The dashed straight line (black), which we show purely for reference,
is equation (\ref{eq:msigagain}) evaluated with a protogalactic
gas fraction of
$f_0\equiv\Omega_{b,0}/(\Omega_{m,0}-\Omega_{b,0})=0.18$ (for the 2013
{\it Planck} cosmology) and the simplistic substitution
$V_{\rm d,pk}\equiv \sqrt{2}\,\sigma_{\rm ap}(R_e)$. The other curves
(blue and red for \citeauthor{dm} and \citeauthor{hern} halo density
profiles) also come from equation (\ref{eq:msigagain}) with
$f_0=0.18$, but with $V_{\rm d,pk}$ depending on
$\sigma_{\rm ap}(R_e)$ as shown in the other panels of Figure
\ref{fig:mvirvpk}. 

The {\it broken} blue and red curves come from those for
$V_{\rm d,pk}$ at $z=0$ versus $\sigma_{\rm ap}(R_e)$ at $z=0$ in the
middle panel of Figure \ref{fig:mvirvpk}. These are the predictions of
equation (\ref{eq:msigagain}) for the critical SMBH masses required to 
clear halos filled with virialised gas in an 18\% mass ratio, via
quasar-mode feedback {\it now}.
It is no surprise that such predictions overshoot the
$M$--$\sigma$ data for normal early-type galaxies, quite
substantially for $\sigma_{\rm ap}(R_e)\ga 200~{\rm km~s}^{-1}$.

The {\it solid} blue and red curves of $M_{\rm BH}$ versus
$\sigma_{\rm ap}(R_e)$, which run through the data, are based on the
curves of $V_{\rm d,pk}$ at $z_{\rm qso}=3$ versus $\sigma_{\rm ap}(R_e)$ at
$z=0$ in the fourth panel of Figure \ref{fig:mvirvpk}. These are
predictions for the $M$--$\sigma$ 
relation in quiescent galaxies at $z=0$, if it came from an
$M_{\rm BH}\propto V_{\rm d,pk}^4$ relationship established by
quasar-mode feedback and blow-out from gaseous protogalaxies at
$z_{\rm qso}=3$ (with negligible subsequent SMBH growth
via coalescence in mergers).

\begin{figure*}
\begin{center}
\includegraphics[width=\textwidth]{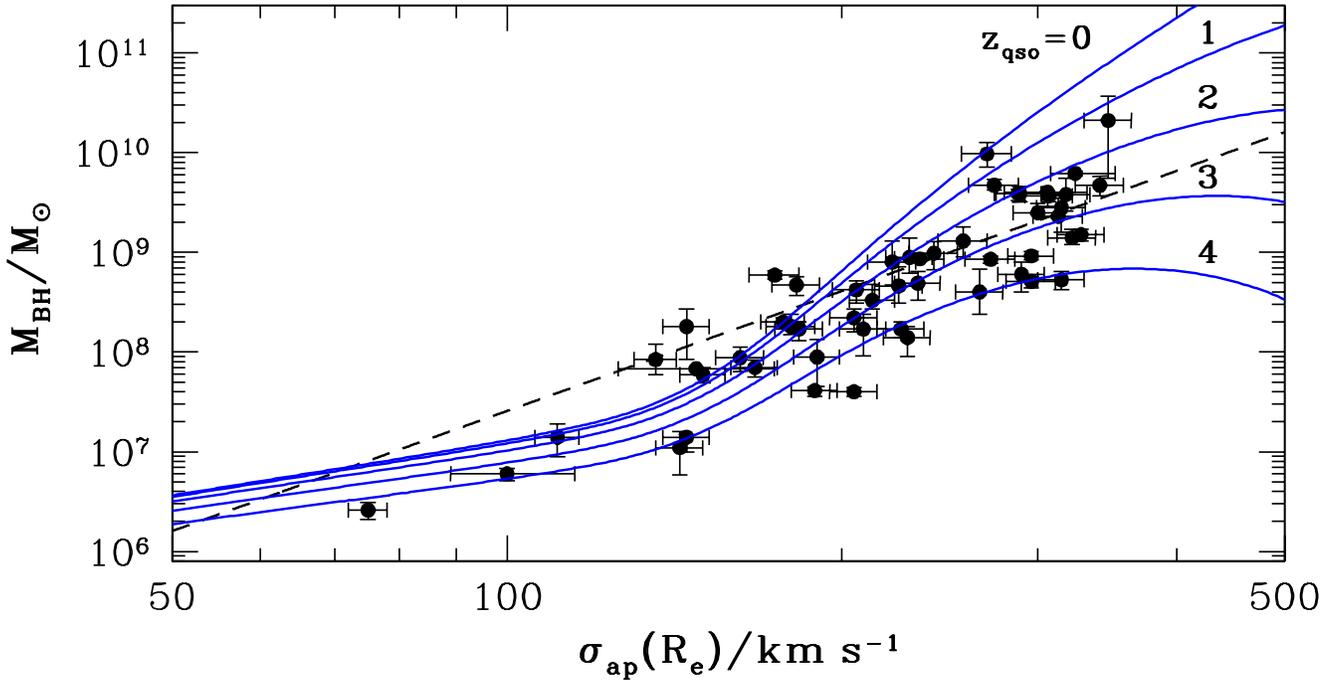}
\end{center} 
\caption{SMBH mass versus stellar velocity dispersion measured inside
  $R_e$ at $z=0$. Data points represent the 53 galaxies flagged as
  early types in \protect\cite{mcconnellma}.
  The solid, blue curves are our models for $M_{\rm BH}$ versus
  $\sigma_{\rm ap}(R_e)$ at $z=0$ if a relation
  $M_{\rm BH}\propto V_{\rm d,pk}^4$ was established by
  accretion-driven feedback,
  according to equation (\ref{eq:msigagain}), at redshift
  $z_{\rm qso}=0,\,1,\,2,\,3~{\rm or}~4$. The curves all assume a 
  \citet{dm} model for the dark-matter halo density profile, and a
  spatially constant gas-to-dark matter mass ratio
  $f_0=0.18$ in the protogalaxies. They do not include any SMBH growth
  between $0<z<z_{\rm qso}$; see text for discussion. For reference
  only, the dashed black line shows equation (\ref{eq:msigagain}) with
  $V_{\rm d,pk}\equiv\sqrt{2}\,\sigma_{\rm ap}(R_e)$.} 
\label{fig:mbhsigma}
\end{figure*}

Figure \ref{fig:mbhsigma} gives an expanded view of $M_{\rm BH}$
versus $\sigma_{\rm ap}(R_e)$. Now, the solid (blue) curves
show SMBH masses obtained from equation (\ref{eq:msigagain}) after
using our scalings to relate stellar velocity dispersion at $z=0$
to the typical $V_{\rm d,pk}$ in progenitor halos at
a wider range of possible $z_{\rm qso}=0,\,1,\,2,\,3~{\rm and}~4$.
All of these curves assume a \citeauthor{dm} density profile for the
dark matter; the results for \citetalias{NFW97} or \citeauthor{hern}
profiles are very similar. The dashed, straight (black) reference line
is again equation (\ref{eq:msigagain}) with
$V_{\rm d,pk}\equiv \sqrt{2}\,\sigma_{\rm ap}(R_e)$.

Most of the $M$--$\sigma$ data at $z=0$ lie between model curves
in which an $M_{\rm BH}$--$V_{\rm d,pk}$ relation emerged from
the clearing of protogalaxies by quasar-mode feedback at redshifts
$2\la z_{\rm qso}\la 4$. The correspondence
of this range with the epoch of peak quasar activity and SMBH
accretion rate in both observations and cosmological simulations
is encouraging.
Equation (\ref{eq:msigagain}) represents a highly simplified,
broad-brush picture of just a few processes at a critical stage of
galaxy and black hole formation; but the fundamental connection it
makes between protogalactic dark-matter halos and SMBH masses
appears to be along the right lines.

The upward bends around
$\sigma_{\rm ap}(R_e)\approx 140~{\rm km~s}^{-1}$ in all of the
$M_{\rm BH}$--$\sigma_{\rm ap}(R_e)$ predictions in Figure
\ref{fig:mbhsigma} trace back to the peak at 
$M_{*,{\rm tot}}\simeq 3.4\times10^{10}\,M_\odot$ [at $z=0$; see
Figure \ref{fig:allplots}(b)] in $f_{*,{\rm vir}}$, the global
stellar-to-dark matter mass fraction. Thus, a linear relation 
$\log(M_{\rm BH})\sim 4\log(V_{\rm d,pk})$ is strongly distorted by a 
non-linear ``conversion'' from halo circular speeds and virial
masses to stellar masses and velocity dispersions.
The curves with $2\le z_{\rm qso}\le 4$ in Figure
\ref{fig:mbhsigma} have average slopes
$\Delta\log M_{\rm BH}/\Delta\log\sigma_{\rm ap}(R_e)\approx 1.5$--2
for galaxies with
$50\la \sigma_{\rm ap}(R_e)\la 100~{\rm km~s}^{-1}$,
but
$\Delta\log M_{\rm BH}/\Delta\log\sigma_{\rm ap}(R_e)\approx 5$--7
in the range
$200\la \sigma_{\rm ap}(R_e)\la 300~{\rm km~s}^{-1}$.
However, this curvature is easily accommodated by the data.
It is reminiscent of the ad hoc, log-quadratic
fits to local $M$--$\sigma$ samples by \citet{wyithe06a,wyithe06b}
(see also \citealt{gultekin09,mcconnellma}).

Equally important is the flattening of the model
$M_{\rm BH}$--$\sigma_{\rm ap}(R_e)$ relations away from the
$z_{\rm qso}=0$ curve, which occurs at high
$\sigma_{\rm ap}(R_e)\ga 300~{\rm km~s}^{-1}$ and
is more pronounced for larger $z_{\rm qso}$.
This is just the behaviour seen in
Figures \ref{fig:vpkz} and \ref{fig:mvirvpk} above:
the masses $M_{\rm d,vir}(z)$ and circular speeds
$V_{\rm d,pk}(z)$ of the most massive 
progenitors of halos (which directly determine $M_{\rm BH}$ here)
have flatter dependences at higher $z$ on the final mass
$M_{\rm d,vir}(0)$ [related to $\sigma_{\rm ap}(R_e)$ at $z=0$ by the
  scalings of \S\ref{sec:modresults}].
Accounting for the generic redshift-evolution of halos in
a $\Lambda$CDM cosmology is critical to the comparison of
models such as equation (\ref{eq:msigagain}) with data at $z=0$.

\subsubsection{Dry mergers at low redshift}
\label{subsubsec:dry}

It is also in the highest-$\sigma_{\rm ap}$
regime that gas-poor galaxy mergers at $z<z_{\rm qso}$ 
may have increased $M_{\rm BH}$ the most from any value determined by
quasar-mode feedback at $z_{\rm qso}$.

\citet{vol13} perform cosmological simulations of black hole growth in
the central galaxies of halos with masses at $z=0$ of
$10^{13}\,M_\odot\le M_{\rm d,vir}(0)\le 10^{15}\,M_\odot$.
They track contributions from gas accretion and from SMBH
coalescences in gas-poor mergers separately.
The results they show for six example halos with
$M_{\rm d,vir}(0)=10^{15}\,M_\odot$ have the central SMBH growth by
accretion essentially finished in all cases at a redshift
$z\approx2$--3. We would associate this here with $z_{\rm qso}$.
Coalescences in dry mergers then drive the growth for $z<z_{\rm qso}$,
and especially at $z\la 1$.
Ultimately the SMBH masses are increased by a wide range of factors,
$f_{\rm co}\equiv M_{\rm BH}(0)/M_{\rm BH}(z_{\rm qso})
   \simeq 1{\mbox{--}}30$.
For a larger sample of $10^{15}$-$M_\odot$ halos, \citeauthor{vol13}
report an average $\langle f_{\rm co}\rangle \approx 11\pm10$.

From \S\ref{sec:modresults}, at $z=0$ the central galaxies in halos
with $M_{\rm d,vir}(0)=10^{15}~M_\odot$ typically have
$M_{*,{\rm tot}}\simeq10^{12}~M_\odot$ and 
$\sigma_{\rm ap}(R_e)\approx 350{\mbox{--}}400~{\rm km~s}^{-1}$
(depending on the assumed dark-matter density profile) .
The rightmost and highest data point in Figure \ref{fig:mbhsigma}
sits near this region; it represents NGC\,4889 in the Coma Cluster,
with $\sigma_{\rm ap}(R_e)=347\pm17~{\rm km~s}^{-1}$
\citep{mcconnell12}. This may well be a system where
low-redshift merging grew $M_{\rm BH}$ substantially above
a feedback-limited value at $z_{\rm qso}=2$--3.

At lower halo and galaxy masses, there is generally much less SMBH
growth through late mergers. For the central galaxies of halos with
$2\times10^{13}\,M_\odot\le M_{\rm d,vir}(0)\le 10^{14}\,M_\odot$
(corresponding to
$M_{*,{\rm tot}}\simeq 2{\mbox{--}}4\times10^{11}\,M_\odot$ and
$\sigma_{\rm ap}(R_e)\approx 220{\mbox{--}}275~{\rm km~s}^{-1}$ at
$z=0$), \citeauthor{vol13} give averages of
$\langle f_{\rm co}\rangle \approx 2\pm1$.
For a set of $10^{13}$-$M_\odot$ halos (corresponding to
$M_{*,{\rm tot}}\simeq1.4\times10^{11}\,M_\odot$ and 
$\sigma_{\rm ap}(R_e)\simeq 200~{\rm km~s}^{-1}$), they find
$\langle f_{\rm co} \rangle=1.8\pm1.8$, suggestive of a small
systematic effect with a few strong outliers.\footnotemark  
\footnotetext{\citeauthor{vol13} do not show explicitly
  for any of their halos with $M_{\rm d,vir}(0)<10^{15}\,M_\odot$ 
  that accretion-driven growth of the central-galaxy SMBH is
  negligible after $z_{\rm qso}\approx2$--3. However, other simulations
  imply this is generally the case (and, indeed, suggest larger
  $z_{\rm qso}$ in some instances); see, e.g., \citet{sijacki07} and
  \citet{dim08}.}

Thus, we can expect dry mergers to scatter data {\it at the top end}
of the $M$--$\sigma$ relation significantly upwards from curves like
those in Figure \ref{fig:mbhsigma}. This would mask any
flattening of the curves at
$\sigma_{\rm ap}(R_e)\ga 300~{\rm km~s}^{-1}$ and
could appear as a much steeper, even near-vertical
mean relation there 
(the so-called ``saturation'' discussed by, e.g.,
\citealt{kho13} and \citealt{mcconnellma}). Among systems with more
moderate velocity dispersions at $z=0$, dry merging can still
introduce some scatter, but not as much. The net
shift up from curves for $M_{\rm BH}$ limited by feedback at
$z_{\rm qso}\simeq2$--3 could plausibly amount to a factor of
$\approx\!2$--3 in the regime
$200\la \sigma_{\rm ap}(R_e)\la 300~{\rm km~s}^{-1}$, and probably
less for lower 
$\sigma_{\rm ap}(R_e)\la 150{\mbox{--}}200~{\rm km~s}^{-1}$.
This should largely preserve the overall shape of such curves.

\subsubsection{Discussion}
\label{subsubsec:msigdiscuss}

Incorporating the generally modest {\it systematic} effects of
low-redshift mergers in the models shown in Figure \ref{fig:mbhsigma}
would primarily move the curves upwards on the plot.
[Mergers at all redshifts are already included in how $V_{\rm d,pk}$
  in a progenitor halo at $z_{\rm qso}>0$ is connected to
  $\sigma_{\rm ap}(R_e)$ in the central galaxy at $z=0$; only the
  value of $M_{\rm BH}$ needs to be adjusted.]
However, a few factors could lower the starting
$M_{\rm BH}$--$V_{\rm d,pk}$ relation predicted by equation
(\ref{eq:msigagain}) at any given $z_{\rm qso}$.

First, if the baryon-to-dark matter mass fraction in a
protogalaxy at $z_{\rm qso}$ were less than
$f_0=0.18$---the cosmic average, assumed for all of the curves in
Figure \ref{fig:mbhsigma}---then the critical $M_{\rm BH}$ for
blow-out would be decreased proportionately.
Second, equation (\ref{eq:msigagain}) ignores any prior work done by a
growing SMBH to push the protogalactic gas outwards before the
point of final blow-out, and thus it overestimates the mass
required to clear the halo completely at $z_{\rm qso}$. Related to
this, lower SMBH masses may suffice to quench quasar-mode accretion by
clearing gas from the inner regions to ``far enough'' away from a
central SMBH, without expelling it fully past the virial
radius.

Cosmological simulations are required to evaluate the balance between
these effects pushing the model $M_{\rm BH}$--$\sigma_{\rm ap}(R_e)$
curves downwards in Figure \ref{fig:mbhsigma}, and the competing
effects of late, dry mergers pulling upwards. But at this level,
the more fundamental simplifications underlying equation
(\ref{eq:msigagain})---among others, the idea that quasar-mode
feedback is always momentum-driven---need to be improved first.

Likewise, low-redshift mergers are just one possible source of 
intrinsic {\it scatter} in the empirical $M$--$\sigma$ relation at
$z=0$. Another is different values in different systems for the
precise redshift at which the main phase of
accretion-driven SMBH growth was ended by quasar-mode
feedback. Even if there were a single $z_{\rm qso}$, there must be real
scatter in the data around any trend line
such as those in Figure \ref{fig:mbhsigma}, because of the scatter
around the constituent scalings from
\S\ref{sec:modingred} and \S\ref{sec:modresults} for halos, halo
evolution and central galaxies.
It is important, but beyond the scope of this paper, to understand the
physical content of the observed $M$--$\sigma$ scatter in detail.
Part of the challenge is to know the ``correct'' trend for
$M_{\rm BH}$ versus $\sigma_{\rm ap}(R_e)$ at $z=0$, around which
scatter should be calculated. In the context of feedback models, this
again requires improving on equation (\ref{eq:msigagain}) for the
prediction of $M_{\rm BH}$ values at $z_{\rm qso}>0$. 

\section{Summary}
\label{sec:summary}

We have examined how a simple relationship between SMBH
masses $M_{\rm BH}$ and the circular speeds $V_{\rm d,pk}$ in
protogalactic dark-matter halos, established by quasar-mode feedback
at redshift $z_{\rm qso}>0$, is reflected in a correlation between
$M_{\rm BH}$ and the stellar velocity dispersions
$\sigma_{\rm ap}(R_e)$ in early-type galaxies at $z=0$.
Straightforward but non-trivial approximations for halo
growth and scalings between halos and their central galaxies
transform a power-law $M_{\rm BH}$--$V_{\rm d,pk}$ relation at
$z_{\rm qso}$ into a decidedly {\it non}--power-law
$M_{\rm BH}$--$\sigma_{\rm ap}(R_e)$ relation at $z=0$. This relation
nevertheless compares well to current data, for assumed 
values of $z_{\rm qso}\approx 2$--4.

We worked with two-component models for spherical galaxies.
Because the stellar properties most relevant to us are those
at (or averaged inside) an effective radius, it sufficed to assume
\cite{hern} density profiles for the stars inside any galaxy. Because
dark-matter halos are key to determining SMBH mass in the feedback
scenario we focussed on, we allowed for any of four different halo
density profiles: those of \citet{NFW96,NFW97}, \citet{hern},
\citet{dm} and \citet{bur95}.

The scaling relations we developed are trend lines connecting
average stellar properties at $z=0$ [total masses $M_{*,{\rm tot}}$,
effective radii $R_e$, aperture velocity dispersions
$\sigma_{\rm ap}(R_e)$ and dark-matter mass fractions] to
the typical virial masses $M_{\rm d,vir}$ and peak circular speeds
$V_{\rm d,pk}$ of dark-matter halos at $z=0$ {\it and} their
most massive progenitors up to $z\la 4$--5. These 
scalings are constrained by theoretical work in the literature on
the global structures, baryon contents and redshift-evolution of
dark-matter halos (\S\ref{sec:modingred}) and by data in the literature
for local elliptical galaxies (\S\ref{sec:modresults}). They 
are robust for normal early-type systems with stellar masses greater
than several $\times10^9\,M_\odot$ at $z=0$, corresponding to velocity
dispersions $\sigma_{\rm ap}(R_e)\ga60{\mbox{--}}70~{\rm km~s}^{-1}$,
but are largely untested against lower-mass dwarf galaxies
(see \S\ref{subsec:discussz0}).

We applied the scalings to show in \S\ref{sec:msigma} how a
relationship of the form $M_{\rm BH}\propto V_{\rm d,pk}^4$
at a range of redshifts $z_{\rm qso}>0$ (equation [\ref{eq:msig}];
\citealt{mcq12}) appears as a much more complicated
$M_{\rm BH}$--$\sigma_{\rm ap}(R_e)$ relation at $z=0$. The specific
form for an initial $M_{\rm BH}$--$V_{\rm d,pk}$ relation comes from
a simplified theoretical analysis of {\it momentum-conserving}
SMBH feedback in {\it isolated and virialised} gaseous
protogalaxies with {\it non-isothermal} dark-matter halos.
Some of the simplifying assumptions involved thus need to be relaxed
and improved in future work. Meanwhile, the highly ``non-linear''
observable $M_{\rm BH}$--$\sigma_{\rm ap}(R_e)$ relation we infer
from it does describe the data for local early types if the redshift
of quasar-mode blow-out was
$z_{\rm qso}\approx 2$--4. This range is reassuringly similar
to the epoch of peak quasar density and SMBH accretion rate in the 
Universe.

This lends support to the notion that the empirical
$M$--$\sigma$ relation fundamentally reflects {\it some} close
connection due to accretion feedback between SMBH masses in
galactic nuclei and the {\it dark matter} in their host (proto)galaxies.
It also demonstrates that the true, physical relationship between
$M_{\rm BH}$ and stellar velocity dispersion at $z=0$ is not
necessarily a pure power law. The shape in our analysis has an
upwards bend around $\sigma_{\rm ap}(R_e)\approx 140~{\rm km~s}^{-1}$
(Figure \ref{fig:mbhsigma}), corresponding to stellar masses
$M_{*,{\rm tot}}\approx3{\mbox{--}}4\times10^{10}\,M_\odot$ and halo 
masses $M_{\rm d,vir}(0)\approx 10^{12}\,M_\odot$ at $z=0$. This bend
comes from a sharp maximum at these masses in the global 
stellar-to-dark matter fractions, $M_{*,{\rm tot}}/M_{\rm d,vir}(0)$
(e.g., \citealt{mos10}). Consequently, there is a sharp upturn in the
dependence of halo circular speeds $V_{\rm d,pk}$ on the stellar
$\sigma_{\rm ap}(R_e)$ (see Figures \ref{fig:allplots} and
\ref{fig:mvirvpk}).

Our models also show a flattening of $M_{\rm BH}$ versus
$\sigma_{\rm ap}(R_e)$ at $z=0$ for velocity dispersions above
$300~{\rm km~s}^{-1}$ or so, for any blow-out redshift $z_{\rm qso}>0$
but more so for higher $z_{\rm qso}$ (Figure \ref{fig:mbhsigma}).
This is due to the way that dark-matter halo masses
grow and circular speeds increase through hierarchical merging in a
$\Lambda$CDM cosmology after $M_{\rm BH}$ is set by feedback and the
halo properties at $z_{\rm qso}$ (see Figure \ref{fig:vpkz}). However,
the values we calculate for $M_{\rm BH}$ include only the growth
by accretion up to $z=z_{\rm qso}$; further growth through SMBH--SMBH
coalescences in gas-poor mergers at lower redshifts is neglected. (The
effects of such mergers on halo masses and circular speeds, and
stellar velocity dispersions at $z=0$, {\it are} accounted for.)

As discussed in \S\ref{subsec:msigma}, simulations by 
\citet{vol13} suggest that low-redshift merging has a significant
effect on the SMBH masses in systems with large
$\sigma_{\rm ap}(R_e)\ga 300$--$350~{\rm km~s}^{-1}$ at $z=0$.
There, dry mergers can scatter $M_{\rm BH}$ values strongly upwards
from the values at $z_{\rm qso}$, essentially erasing the  
flattening that might otherwise be observed at $z=0$ and
``saturating'' the empirical $M$--$\sigma$ relation. In galaxies with
lower $\sigma_{\rm ap}(R_e)\la 300~{\rm km~s}^{-1}$, where most
current data fall, such scatter up from feedback-limited SMBH
masses will be much more modest in general. The expected
$M_{\rm BH}$--$\sigma_{\rm ap}(R_e)$ relations at $z=0$
should then have the same basic shape as when late mergers are
ignored.

Although we have focussed on the observed $M$--$\sigma$ relation,
other SMBH--bulge correlations exist that may be just as strong
intrinsically. These include the $M_{\rm BH}{\mbox{--}}M_{\rm bulge}$
correlation and multivariate, ``fundamental-plane'' relationships
between $M_{\rm BH}$ and non-trivial combinations of $M_{*,{\rm tot}}$,
$R_e$ and $\sigma_{\rm ap}(R_e)$. They should also reflect any
underlying SMBH--dark matter connection at some $z_{\rm qso}>0$, and
the techniques of this paper can be applied to
look at them as well. However, this will best be done
with close attention also paid to the inevitable
scatter around all of the scalings we have adopted for both stellar
and  dark-matter halo properties. It remains to be
understood how the numerous individual sources
of scatter combine to produce SMBH correlations with apparently so
little net scatter at $z=0$.

More sophisticated predictions of critical SMBH masses for quasar-mode
blow-out in terms of protogalactic dark-matter halo properties are
required. The simple relation $M_{\rm BH}\propto V_{\rm d,pk}^4$ that
we have used makes very specific assumptions about the mechanism
(e.g., momentum-driven) and the setting (spherical protogalaxies with
no stars, initially virialised gas, smooth outflows) of the feedback
that establishes it. We mentioned in \S\ref{subsec:mbhvhalo} and
\S\ref{subsec:msigma} several ways to improve on these assumptions.
Our work in this paper is readily adaptable to help test any refinements.

\section*{Acknowledgements}
ACL has been supported by an STFC studentship.

\appendix

\section{Model checks at \,{\boldmath$\MakeLowercase{z}=0$}}
\label{sec:appendix}

Here we collect some properties from the literature for a few
galaxies and halos spanning the range of mass and stellar velocity
dispersion covered by local galaxy samples used to define
empirical SMBH $M$--$\sigma$ relations. We then extract numerical
values from the $z=0$ scalings in \S\ref{sec:modresults}
(Figure \ref{fig:allplots}) to compare with the measurements.

\subsection{Stellar and halo properties from the literature}
\label{subsec:measun}

Table \ref{tab:measure_table} lists observed stellar properties 
of the Milky Way, M87 (at the centre of Virgo subcluster A), M49 (at
the centre of Virgo~B) and NGC\,4889 (in the Coma
Cluster). Properties of the dark matter halos are also given, from
dynamical modelling in the literature. Our analysis is clearly not
meant to describe disc galaxies, but we have included the Milky Way
as a useful check on the implications for $\sim\!L^\star$
galaxies in general.

\subsubsection{The Milky Way}

In the first row of  Table \ref{tab:measure_table}, the total stellar
mass, the radius $r_{\rm 200}$ of mean overdensity $\Delta\equiv 200$
and the dark-matter mass $M_{{\rm d},200}$ inside this  are all taken
from \cite{mcmillan11}. Combining his best-fitting
\citetalias{NFW97} concentration,
$r_{200}/r_{-2}\simeq 9.55$, with his values of $M_{\rm{d,200}}$ and
$r_{200}$ plus $r_{\rm pk}/r_{-2}=2.16258$ for an
\citetalias{NFW97} halo, yields $r_{\rm{pk}}\simeq 52\, \rm{kpc}$ and
$V_{\rm{d,pk}}\simeq 185\, \rm{km\,s^{-1}}$. These are consistent with
separate modelling of the Milky Way by \citet{dms06}. 

The second row of Table \ref{tab:measure_table} contains the total
stellar mass of the Milky Way bulge only, according to
\citet{mcmillan11}. He does not record the effective radius of the
bulge or the aperture dispersion inside it, so we take
$R_e\simeq 2.7~\rm{kpc}$ from \cite{free85} and
$\sigma_{\rm{ap}}(R_e)\simeq 103~{\rm km~s}^{-1}$ from
\cite{mcconnellma}.

\subsubsection{M87 and M49}

For M87 and M49, Table \ref{tab:measure_table} quotes total stellar
masses based on three different sources:
the ATLAS$^{\rm 3D}$ survey \citep{capI}, the
ACSVCS \citep{chen} and \citet{mcconnellma}.
The original authors give total luminosities, to which we have
applied mass-to-light ratios from \citet{Mar05} models for a
\citet{kroupa01} IMF and a stellar age of 9~Gyr:
$M_{*,{\rm tot}}/L_K\simeq 0.88~M_\odot\,L_\odot^{-1}$ for the
ATLAS$^{\rm 3D}$ luminosity, 
$M_{*,{\rm tot}}/L_z\simeq 1.7~M_\odot\,L_\odot^{-1}$ for the ACSVCS
value and
$M_{*,{\rm tot}}/L_V\approx 3.15~M_\odot\,L_\odot^{-1}$ for
\citet{mcconnellma}.
Both galaxies have $R_e$ values in the ATLAS$^{\rm 3D}$ 
survey and the ACSVCS, and velocity dispersions in ATLAS and
\cite{mcconnellma}.

\citet{mcl99} and \citet{cote01} fitted 
the kinematics of stars and globular clusters in M87, plus the
kinematics of Virgo-cluster galaxies and the total mass
profile derived from intracluster X-ray gas, with a
two-component mass model comprising the stars (plus remnants and
stellar ejecta) in the body of M87 and an \citetalias{NFW97}
dark-matter halo with $r_{\rm 200}\simeq 1.55$~Mpc and
$M_{\rm d,200}\simeq 4.2\times10^{14}~M_\odot$. This clearly
identifies the dark matter in and around M87 with the halo of
the entire Virgo~A subcluster. \citeauthor{mcl99} and
\citeauthor{cote01} have an \citetalias{NFW97}
concentration of $r_{\rm 200}/r_{-2}=2.8\pm0.7$
for the M87/Virgo~A halo, so (with $r_{\rm pk}/r_{-2}=2.16258$ again)
$r_{\rm pk}\sim 1.2$~Mpc and
$V_{\rm d,pk}\simeq 1100~{\rm km~s}^{-1}$.

For M49/Virgo~B, \citet{cote03} similarly use a two-component mass
model consisting of the galaxy's stars plus a single
\citetalias{NFW97} dark-matter
halo, to fit the stellar and globular cluster kinematics on
$\la\!50$-kpc scales and the X-ray mass profile
out to $\sim\!{\rm Mpc}$ radii. The \citeauthor{cote03} analysis
implies $r_{200}\simeq 950~{\rm kpc}$ with
$M_{\rm d,200}\simeq 9.4\times10^{13}~M_\odot$, and
$r_{\rm 200}/r_{-2}\simeq 4.8$. The dark-matter circular speed
therefore peaks at $r_{\rm pk}\simeq 425~{\rm kpc}$, where
$V_{\rm d,pk}\simeq 710~{\rm km~s}^{-1}$.


\begin{table*}
\centering
\caption{Values of stellar and dark matter halo properties at
          $z=0$, taken from various sources in the
          literature.
          References: 1 -- \citet{mcmillan11}, 2 -- \citet{free85}, 
           3 -- \citet{mcconnellma}, 4 -- \citet{capI}, 
           5 -- \citet{capxv},
           6 -- \citet{mcl99}, 7 -- \citet{cote01},
           8 -- \citet{chen}, 9 -- \citet{cote03}, 
           10 -- \citet{mcconnell11,mcconnell12},
           11 -- \citet{lokas03}. }
\label{tab:measure_table}
\begin{tabular}{ lcccccccccc }
        \hline

	Galaxy & $M_{\rm{*,tot}}$ & $R_e$ & ref. &
                 $\sigma_{\rm{ap}}(R_e)$ & ref. & $V_{\rm{d,pk}}$ &
                 $r_{\rm{pk}}$ & $M_{\rm{d,200}}$ or $M_{\rm d,vir}$ &
                 $r_{\rm{200}}$ or $r_{\rm vir}$&
                 ref. \\ 
               & $(M_\odot)$ & (kpc) & & (km~s$^{-1}$) & &
                 (km~s$^{-1}$) & (kpc) & $(M_\odot$) & (kpc) \\
        \hline

        Milky Way & $6.4\times 10^{10}$ & -- & 1 & -- & -- & 185 &
              52 & $1.26\times 10^{12}$ & 230 & 1\\ 

        MW bulge & $9.0\times 10^{9~}$ & $2.7\pm 0.3$ & 1,2 &
              $103\pm 20$ & 3 & -- & -- & -- & -- & -- \\ 

        \hline

        M87 & $2.9\times 10^{11}$ & $6.8\pm 1.5$ & 4 & $264\pm 13$ &
              5 & 1100 & 1200 & $4.2\times10^{14}$ & 1550 & 6,7 \\ 
            & $3.2\times 10^{11}$ & $8.7\pm 1.1$ & 8 & -- & -- &
              -- & -- & -- & -- & -- \\ 
            & $3.7\times 10^{11}$ & -- & 3 & $324^{+28}_{-16}$ & 3 &
              -- & -- & -- & -- & -- \\ 

        \hline

  	M49 & $4.2\times 10^{11}$ & $7.9\pm 1.7$ & 4 & $250\pm13$ &
              5 & 710 & 425 & $9.4\times 10^{13}$ & 950 & 9 \\ 
            & $4.7\times 10^{11}$ & $13.4\pm 1.1$ & 8 & -- & -- & -- &
              -- & -- & -- & -- \\ 
            & $3.7\times 10^{11}$ & -- & 3 & $300\pm 15$ & 3 &	-- &
              -- & -- & -- & -- \\ 

        \hline

	NGC\,4889 & $9.5\times 10^{11}$ & $27\pm 2$ & 3,10 &
                  $347\pm 17$ & 3,10 & 1585 & 670 &
                  $1.2\times 10^{15}$ & 2900 & 11 \\

        \hline      
\end{tabular}
\end{table*}


\begin{table*}
\centering
\caption{Stellar and dark matter halo properties at $z=0$ according to
  our scaling relations. For each galaxy, the starting point is
  $M_{\rm{*,tot}}$, taken from the literature.}
\label{tab:compare_table}
\begin{tabular}{ lccccccc }
	\hline
                Galaxy & 
                $M_{\rm{*,tot}}$ & 
                $R_e$  &                 
                $\sigma_{\rm{ap}}(R_e)$ &                 
                $V_{\rm{d,pk}}$ &               
                $r_{\rm{pk}}$ &                
                $M_{\rm{d,200}}$ or $M_{\rm d,vir}$ &          
                $r_{\rm{200}}$ or $r_{\rm vir}$ \\
           		  & 
		  $(M_\odot)$ & 
		  (kpc) &	 
		  (km~s$^{-1}$) & 
		  (km~s$^{-1}$) & 
		  (kpc) & 
		  ($M_\odot$) & 
		  (kpc) \\
	\hline  

        Milky Way & $6.4\times 10^{10}$ & ~\,3.0 & 160   & ~\,200 &
            ~\,75 & $2.0\times 10^{12}$ & ~\,270 \\ 

        MW bulge  & $9.0\times 10^{9~}$ & ~\,1.4 & ~\,90 & ~\,120 &
            ~\,35 & $3.6\times 10^{11}$ & ~\,150 \\ 

        M87       & $3.3\times 10^{11}$ & ~\,8.0 & 245   & ~\,600 &
              330 & $6.0\times 10^{13}$ & ~830 \\ 

        M49       & $4.2\times 10^{11}$ & ~\,9.3 & 265   & ~\,720 &
              420 & $1.0\times 10^{14}$ & 1000 \\ 

        NGC\,4889 & $9.5\times 10^{11}$ & 15.2   & 345   &   1285 &
              925 & $8.0\times 10^{14}$ & 2450 \\ 

	\hline
\end{tabular}
\end{table*}


\subsubsection{NGC 4889}

NGC\,4889 is the brightest galaxy in Coma and not far from the
nominal central galaxy, NGC\,4874. According to
\citet{mcconnellma},
NGC\,4889 has $L_V\simeq 3.0\times 10^{11}L_\odot$
and hence (for $M_*/L_V \approx 3.15~M_\odot\,L_\odot^{-1}$ 
from the \citealt{Mar05} population-synthesis models)
$M_{*,{\rm tot}}\approx 9.5\times10^{11}~M_\odot$. It is at
the uppermost end of the range of stellar masses plotted for
our relations in Figure \ref{fig:allplots} (but it does not appear
on those plots since it is not in the ATLAS$^{\rm 3D}$
survey), and it hosts one of the largest supermassive black
holes yet measured: $M_{\rm BH}=(2.1\pm1.6)\times10^{10}~M_\odot$
\citep{mcconnell11,mcconnell12}. The effective radius $R_e=27$~kpc and
velocity dispersion $\sigma_{\rm ap}(R_e)=347~{\rm km~s}^{-1}$ in Table
\ref{tab:measure_table} are from
\cite{mcconnellma} and \citet{mcconnell11,mcconnell12}. 

The global dark matter properties of the Coma Cluster are taken from
dynamical modelling by \cite{lokas03}. They give values for 
$r_{\rm{vir}}$ and $M_{\rm{d,vir}}$, rather than $r_{200}$ and
$M_{\rm{d,200}}$ like the other galaxies in
Table \ref{tab:measure_table}, and a 
best-fitting \citetalias{NFW97} concentration of
$r_{\rm vir}/r_{-2}=9.4$. Together these imply 
$r_{\rm pk}\simeq 670~{\rm kpc}$ and
$V_{\rm d,pk}\simeq 1585~{\rm km~s}^{-1}$.

\subsection{Comparison to models} 

Taking the total stellar mass $M_{\rm{*,tot}}$ as a starting point
for each of the systems in Table \ref{tab:measure_table}, we now find
their other stellar and halo properties from the scaling relations
developed in \S\ref{sec:modresults}.
Table \ref{tab:compare_table} shows the results for
$R_e$, $\sigma_{\rm{ap}}(R_e)$, $V_{\rm{d,pk}}$, $r_{\rm{pk}}$,
$M_{\rm{d,200}}$ or (for NGC\,4889/Coma) $M_{\rm d,vir}$, and
$r_{\rm{200}}$ or (for NGC\,4889/Coma) $r_{\rm vir}$.  

\subsubsection{$L^{\star}$ galaxies:\,
       $\sigma_{\rm ap}(R_e) \sim 100$--$150~{\rm km~s}^{-1}$}

For $M_{\rm{*,tot}}\simeq 6.4\times10^{10}M_\odot$ (the {\it total} Milky
Way mass), our scalings give the stellar effective radius as
$R_e\simeq 3~{\rm kpc}$ and the velocity dispersion as
$\sigma_{\rm ap}(R_e)\simeq 160~{\rm km~s}^{-1}$. 
This dispersion is rather higher than the value typically used to put
the Milky Way on the black hole $M$--$\sigma$ relation: for
example, \citet{mcconnellma} take
$\sigma_{\rm ap}(R_e)=103~{\rm km~s}^{-1}$ for the Galaxy.
However, this value is meant to represent the bulge only.
For the {\it bulge} mass of 
$M_{*,{\rm tot}}\simeq 9\times10^{9}~M_\odot$, our relations
give $R_e\simeq 1.4$~kpc and
$\sigma_{\rm ap}(R_e)\simeq 90~{\rm km~s}^{-1}$.

For the total Galactic stellar mass of $6.4\times10^{10}M_\odot$ and
assuming an \citetalias{NFW97} halo, the scalings lead to
a peak circular speed of $V_{\rm d,pk}\simeq 200~{\rm km~s}^{-1}$,
occurring at $r_{\rm pk}\simeq 75$~kpc. 
Using equations (\ref{eq:nfwmass}), (\ref{eq:dut}) and (\ref{eq:mrvir})
to go from the virial radius implied by $M_{*,{\rm tot}}$ to
the radius of mean overdensity $\Delta=200$, we find
$M_{\rm{d,200}}\simeq 2\times 10^{12}M_\odot$ and
$r_{200}\simeq 270$ kpc.
For the mass of the bulge alone,
$M_{\rm{*,tot}}\simeq 9\times 10^9M_\odot$, 
we obtain $V_{\rm{d,pk}}\sim 120\, \rm{km\,s^{-1}}$,
$r_{\rm{pk}}\sim 35$ kpc, $M_{\rm{d,200}}\sim 3.6\times 10^{11}M_\odot$
and $r_{200}\sim 150$ kpc.
  
\subsubsection{M87 and M49:\,
       $\sigma_{\rm ap}(R_e) \sim 250~{\rm km~s}^{-1}$}
  
For each of these galaxies, we take the mean of $M_{*,{\rm tot}}$ from
the three different values in Table \ref{tab:measure_table}. Thus,
$M_{*,{\rm tot}}=3.3\times10^{11}~M_\odot$ for M87, and 
$M_{\rm{*,tot}}=4.2\times10^{11}~M_\odot$ for M49.
Our parametrisation of $R_e$ versus $M_{*,{\rm tot}}$ in
\S\ref{subsec:re} then gives the values recorded in Table
\ref{tab:compare_table}, which broadly agree with the measurements of
$R_e$. The model values in Table \ref{tab:compare_table} for
$\sigma_{\rm{ap}}(R_e)$, $V_{\rm{d,pk}}$, $r_{\rm{pk}}$,
$M_{\rm{d,200}}$ and $r_{\rm{200}}$ assume an NFW halo around each
galaxy (as the analyses from the literature do). The predicted
velocity dispersions compare well to the measurements for M87 and M49
in the ATLAS$^{\rm 3D}$ survey but not quite as well to the
values recorded by \citet{mcconnellma}, which are $\!20\%$ higher.

The value of $r_{200}$ for M87/Virgo~A in Table
\ref{tab:measure_table}, from \citet{mcl99}, is $\simeq\!80\%$  bigger
than the one in Table \ref{tab:compare_table}, implied by our models
here. \citeauthor{mcl99}'s $M_{\rm d,200}$ is consequently larger by
about a factor of $1.8^3\simeq 6$. Similarly,
the circular-speed curve of the halo in \citet{mcl99} peaks at
$r_{\rm pk}\sim 1.2$~Mpc (with a very large uncertainty) rather than
$r_{\rm pk}\simeq 330$~kpc as expected here, and it has
$V_{\rm d,pk}\simeq 1100~{\rm km~s}^{-1}$ rather than
$V_{\rm d,pk}\simeq 600~{\rm km~s}^{-1}$.

These discrepancies for M87/Virgo~A may simply reflect the inevitable
scatter in the properties of individual systems around the typical
values given by our trend lines. For M49/Virgo~B, all of the halo
properties in Table \ref{tab:compare_table} obtained from our scalings
are remarkably close to the values in Table
\ref{tab:measure_table} from \citet{cote03}.

\subsubsection{NGC\,4889:\,
       $\sigma_{\rm ap}(R_e) \sim 350~{\rm km~s}^{-1}$}

For $M_{*,{\rm tot}}=9.5\times10^{11}~M_\odot$, our scalings give
$R_e=15.2~{\rm kpc}$ and (assuming an \citetalias{NFW97} halo)
$\sigma_{\rm ap}(R_e)\simeq 345~{\rm km~s}^{-1}$. 
The velocity dispersion agrees with the value in
\citet{mcconnell11,mcconnell12}, although the effective radius is
smaller than their adopted 27~kpc. Further, we find
$r_{\rm vir}\simeq 2.45~{\rm Mpc}$ and
$M_{\rm d,vir}\simeq 8.0\times10^{14}~M_\odot$, which compare well to
the values in Table \ref{tab:measure_table} determined by
\cite{lokas03}. (This is even though NGC\,4889 is not precisely at the
centre of the Coma Cluster).

Assuming an \citetalias{NFW97} halo density profile, our models
imply $r_{\rm pk}\simeq 925~{\rm km~s}^{-1}$ and
$V_{\rm d,pk}\simeq 1285~{\rm km~s}^{-1}$ for the peak of the
dark-matter circular speed in NGC\,4889/Coma---different by
$\sim\!30\%$ from the \citeauthor{lokas03} numbers.
Comparing to the peak radii and speeds above for M87/Virgo~A and
M49/Virgo~B emphasises the clear visual impression given by
Figure \ref{fig:allplots}:
In large galaxies $V_{\rm d,pk}$, along with $M_{\rm d,vir}$, is a
much more sensitive function of galaxy stellar mass than the 
stellar $\sigma_{\rm ap}(R_e)$ is.
(This follows directly from the steep decline at high masses in the
cosmological connection between $M_{*,{\rm tot}}$ and $M_{\rm d,vir}$
adopted from \citealt{mos10}.)
It therefore seems natural to expect much more scatter and many more
apparent ``outliers'' in $M_{\rm BH}$ among very massive galaxies, if
SMBH masses are connected fundamentally to the 
global properties of dark-matter halos rather than to stellar velocity 
dispersions directly. 

\label{lastpage}
\end{document}